\documentclass[twocolumn,notitlepage,prl,superscriptaddress,longbibliography]{revtex4-2}

\usepackage{amsfonts}
\usepackage{textcomp}
\usepackage{times}
\usepackage{graphicx}
\usepackage{float}
\usepackage{latexsym,amsmath,amssymb,bm,euscript}
\usepackage{color}
\usepackage{subfigure}
\usepackage{epstopdf}
\usepackage[colorlinks=true,linkcolor=blue,citecolor=blue]{hyperref}
\usepackage{soul}
\usepackage[normalem]{ulem}
\usepackage{mathrsfs}
\usepackage{amsmath}
\usepackage{xspace}
\usepackage{natbib}
\usepackage{ulem}
\usepackage{etoolbox}
\usepackage{tikz}
\usepackage{physics}
\usepackage{chemformula}
\usepackage{natbib}
\usepackage{tikz}

\newcommand{\ttrg}{{\emph{tan}TRG}}

\newcommand{\gr}{Gr\"uneisen ratio}
\newcommand{\scbo}{\ch{SrCu_2(BO_3)_2}}
\newcommand{\ZZ}{\mathbb{Z}_2}

\begin{document}
\title{Plaquette Singlet Transition, Magnetic Barocaloric Effect, and Spin Supersolidity in the Shastry-Sutherland Model}

\author{Junsen Wang}
\thanks{These authors contributed equally to this work.}
\affiliation{Center of Materials Science and Optoelectronics Engineering,
College of Materials Science and Opto-electronic Technology, University
of Chinese Academy of Sciences, Beijing 100049, China.}
\affiliation{CAS Key Laboratory of Theoretical Physics, Institute of Theoretical
Physics, Chinese Academy of Sciences, Beijing 100190, China}

\author{Han Li}
\thanks{These authors contributed equally to this work.}
\affiliation{Kavli Institute for Theoretical Sciences, University of Chinese Academy of Sciences, Beijing 100190, China}
\affiliation{CAS Key Laboratory of Theoretical Physics, Institute of
Theoretical Physics, Chinese Academy of Sciences, Beijing 100190, China}

\author{Ning Xi}
\affiliation{CAS Key Laboratory of Theoretical Physics, Institute of Theoretical
Physics, Chinese Academy of Sciences, Beijing 100190, China}

\author{Yuan Gao}
\affiliation{School of Physics, Beihang University, Beijing 100191, China}
\affiliation{CAS Key Laboratory of Theoretical Physics, Institute of
Theoretical Physics, Chinese Academy of Sciences, Beijing 100190, China}

\author{Qing-Bo Yan}
\email{yan@ucas.ac.cn}
\affiliation{Center of Materials Science and Optoelectronics Engineering,
College of Materials Science and Opto-electronic Technology, University
of Chinese Academy of Sciences, Beijing 100049, China.}

\author{Wei Li}
\email{w.li@itp.ac.cn}
\affiliation{CAS Key Laboratory of Theoretical Physics, Institute of
Theoretical Physics, Chinese Academy of Sciences, Beijing 100190, China}
\affiliation{CAS Center for Excellence in Topological Quantum Computation,
University of Chinese Academy of Sciences, Beijing 100190, China}
\affiliation{Hefei National Laboratory, University of Science and Technology of China, Hefei 230088, China}

\author{Gang Su}
\email{gsu@ucas.ac.cn}
\affiliation{Kavli Institute for Theoretical Sciences, University of Chinese Academy of Sciences, Beijing 100190, China}
\affiliation{CAS Center for Excellence in Topological Quantum Computation,
University of Chinese Academy of Sciences, Beijing 100190, China}

\date{\today}

\begin{abstract}
Inspired by recent experimental measurements [Guo \textit{et al.}, 
Phys. Rev. Lett.~\textbf{124}, 206602 (2020); Jiménez \textit{et al.}, 
Nature \textbf{592}, 370 (2021)] on frustrated quantum magnet {\scbo} 
under combined pressure and magnetic fields, we study the related 
spin-$1/2$ Shastry-Sutherland (SS) model using state-of-the-art tensor 
network methods. By calculating thermodynamics, correlations and 
susceptibilities, we find, in zero magnetic field, not only a line of first-order 
dimer-singlet to plaquette-singlet (PS) phase transition ending with a 
critical point, but also signatures of the ordered PS transition with its 
critical endpoint terminating on this first-order line. Moreover, we uncover
prominent magnetic barocaloric responses, a novel type of quantum 
correlation induced cooling effect, in the strongly fluctuating supercritical 
regime. Under finite fields, we identify a quantum phase transition from 
the PS phase to the spin supersolid phase that breaks simultaneously 
lattice translational and spin rotational symmetries. The present findings 
on the SS model are accessible in current experiments and would shed 
new light on the critical and supercritical phenomena in the archetypal 
frustrated quantum magnet {\scbo}.
\end{abstract}

\maketitle

\emph{Introduction.---}
Frustrated magnetism constitutes a fertile ground breeding enriched 
spin states and phase transitions~\cite{lacroix2011,diep2020}, including 
unusual spin orders, quantum spin liquid (QSL)~\cite{balents2010,
zhou2017,broholm2020}, and unconventional quantum critical point 
(QCP) like the deconfined QCP (DQCP)~\cite{senthil2004}, etc. The 
paradigmatic Shastry-Sutherland (SS) model is a highly frustrated 
quantum spin system with an analytically known ground state in certain 
parameter regime~\cite{shastry1981}. 
Nevertheless, its global phase diagram hosts rich spin states and 
transitions, where numerical simulations are playing an increasingly 
important role~\cite{albrecht1996,miyahara1999,
muller2000,koga2000,zheng2001,takushima2001,chung2001,lauchli2002,
isacsson2006,lou2012,corboz2013,boos2019,lee2019,shimokawa2021,
xi2023,keles2022,wang2022,yang2022}. On the other hand and
as a miracle of nature, the SS model is faithfully realized by a quantum 
magnetic material {\scbo} whose pressure-field-temperature phase 
diagram is under intensive investigation~\cite{kageyama1999,radtke2015,
zayed2017,mcclarty2017,guo2020,jimnez2021,cui2022,shi2022,
nomura2022}, and the intriguing magnetic phenomena observed in 
experiments in turn require further theoretical studies of the SS model.

\begin{figure}
\centering
\includegraphics[width=\linewidth]{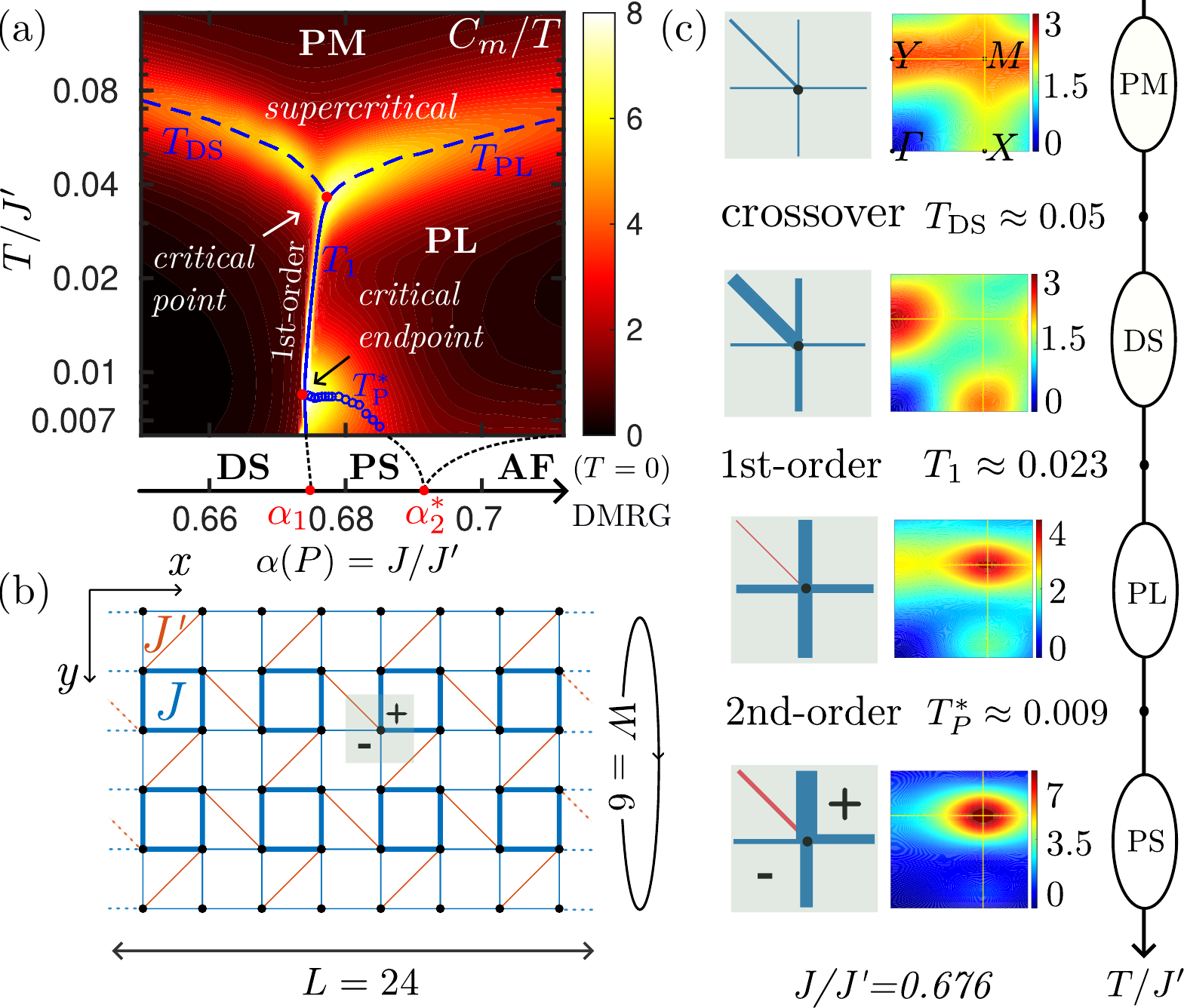}
\caption{(a) Pressure-temperature phase diagram of the SS model 
with magnetic specific heat $C_m/T$ as contour background. The 
low-temperature data smoothly extrapolate to the ground-state results,
where a first-order QPT occurring at $\alpha_1\simeq 0.673$ and a 
QCP at $\alpha_2^\ast \simeq 0.692$ are obtained by DMRG~\cite{SM} 
on the same geometry. Two crossovers $T_{\rm DS}$ and $T_{\rm PL}$ 
(blue dashed lines), first-order $T_1$ (solid line), and second-order 
transition $T_{\rm P}^*$ (empty circle for each data point) are 
determined from peaks of $C_m/T$.
(b) The 6$\times$24 cylinder with the ordered PS phase illustrated. 
(c) Temperature evolution of local bond correlators (first column), static
spin structure factors (second column) and the corresponding phases
(third column) for $J/J'=0.676$. Local correlators are measured at the 
center of the lattice, exemplified by the light green square in (b), with two 
types of empty plaquettes indicated by $+$ and $-$ signs, respectively. 
Blue (red) bonds indicate negative (positive) correlations with their 
widths proportional to the absolute values.}
\label{fig1}
\end{figure}

The spin-1/2 SS model is defined on a square lattice with the Hamiltonian
\begin{equation}
    H= J\sum_{\expval{i,j}} \vb S_i \cdot \vb S_j +
    J'\sum_{\expval{\expval{i,j}}}\vb S_i \cdot \vb S_j,
    \label{eq:h}
\end{equation}
where $J>0$ is the antiferromagnetic (AF) coupling on the inter-dimer, 
and $J'>0$ on the intra-dimer bonds [c.f., Fig.~\ref{fig1}(b)]. We take 
$J'=1$ as the energy scale hereafter. For $\alpha \equiv J/J' \leq 0.5$, 
the ground state is rigorously 
a product of singlets on the $J'$ dimers~\cite{shastry1981}, dubbed the 
dimer-singlet (DS) phase. While for the other limit, $\alpha\gg 1$, it has 
clearly a N\'eel AF state~\cite{manousakis1991}, and possible intermediate 
phases were debated for decades~\cite{albrecht1996,miyahara1999,
muller2000,koga2000,zheng2001,takushima2001,chung2001,lauchli2002,
isacsson2006,lou2012}. Now consensus has more or less reached that there
exists an intervening plaquette-singlet (PS) phase~\cite{corboz2013,zayed2017,
boos2019,lee2019,guo2020,shimokawa2021,xi2023,cui2022,keles2022,shi2022,
wang2022,yang2022}.
The DS-PS transition is first-order, while the PS-AF transition is possibly
second-order and belongs to a DQCP~\cite{lee2019}. Evidence of the 
intermediate PS phase and pressure-induced quantum phase transitions (QPTs)
were indeed found in recent experiments~\cite{zayed2017,guo2020,jimnez2021}

However, there are still enigmas surrounding this seemingly innocent
PS phase. For example, the experiments on {\scbo} show that the 
plaquette singlets sit on ``full'' plaquettes containing diagonal bonds
\cite{bettler2020,zayed2017,cui2022}; while recent ground-state 
numerics suggest the ``empty'' plaquettes without diagonal bonds 
instead~\cite{corboz2013,xi2023,lee2019,yang2022,wang2022,keles2022,shimokawa2021}. As a crucial step towards resolving 
this discrepancy, a finite-temperature calculation from the theoretical 
side is much in demand first: How about the competition between 
the instabilities towards empty and full PS order \emph{across the full 
temperature range}? After all, based only on the ground-state 
results one cannot exclude in principle a full-plaquette phase at an 
intermediate temperature. Recent specific heat measurements on 
{\scbo} found a signal at $T\sim 2$~K, conjecturing that it reflects 
the onset of PS order~\cite{guo2020,jimnez2021}. Despite of great 
efforts for thermodynamics simulations made in recent years~\cite{wessel2018,
wietek2019,shimokawa2021}, this signal remains elusive theoretically
due to the great challenges in unbiased calculations down to such low temperature.

In this work, we perform a finite-temperature study of the SS model
on a cylinder geometry with the state-of-the-art exponential tensor 
renormalization group (XTRG) approach~\cite{Chen2017,Chen2018,
Lih2019}. XTRG has been successfully used in studying frustrated 
quantum magnets~\cite{Li2020TMGO,Li2021RuCl,Yu2021,Gao2022NBCP}, 
and here we simulate the SS model down to $T/J'\sim 0.006$ on a 
long cylinder. By mapping out the phase diagram, we reproduce the 
critical point~\cite{jimnez2021} and uncover the empty PS phase below 
the thermal transition line $T_{\rm P}^*$ with a $\mathbb{Z}_2$ symmetry 
breaking. Although the calculations are restricted within $W=6$ cylinders, 
we believe the conclusions also hold for wider ones (preliminary 
width-8 results also support this scenario, see Supplemental Fig.~S2~\cite{SM}), and therefore explain the specific heat peak 
observed at $T \lesssim 2$~K in recent experiments~\cite{guo2020,
jimnez2021}. We further propose a pronounced quantum 
correlation cooling driven by pressure (that controls the coupling ratio 
$\alpha$) in the supercritical regime. As a companion, we also perform 
density-matrix renormalization group (DMRG)~\cite{white1992} calculations 
to explore the QPTs driven by combined pressure and magnetic fields. 
In particular, evidence for the QPT between the PS phase and a spin 
supersolid (SSS) phase is witnessed, whose location well 
agrees with recent experiments~\cite{cui2022}.

\begin{figure}
\centering
\includegraphics[width=\linewidth]{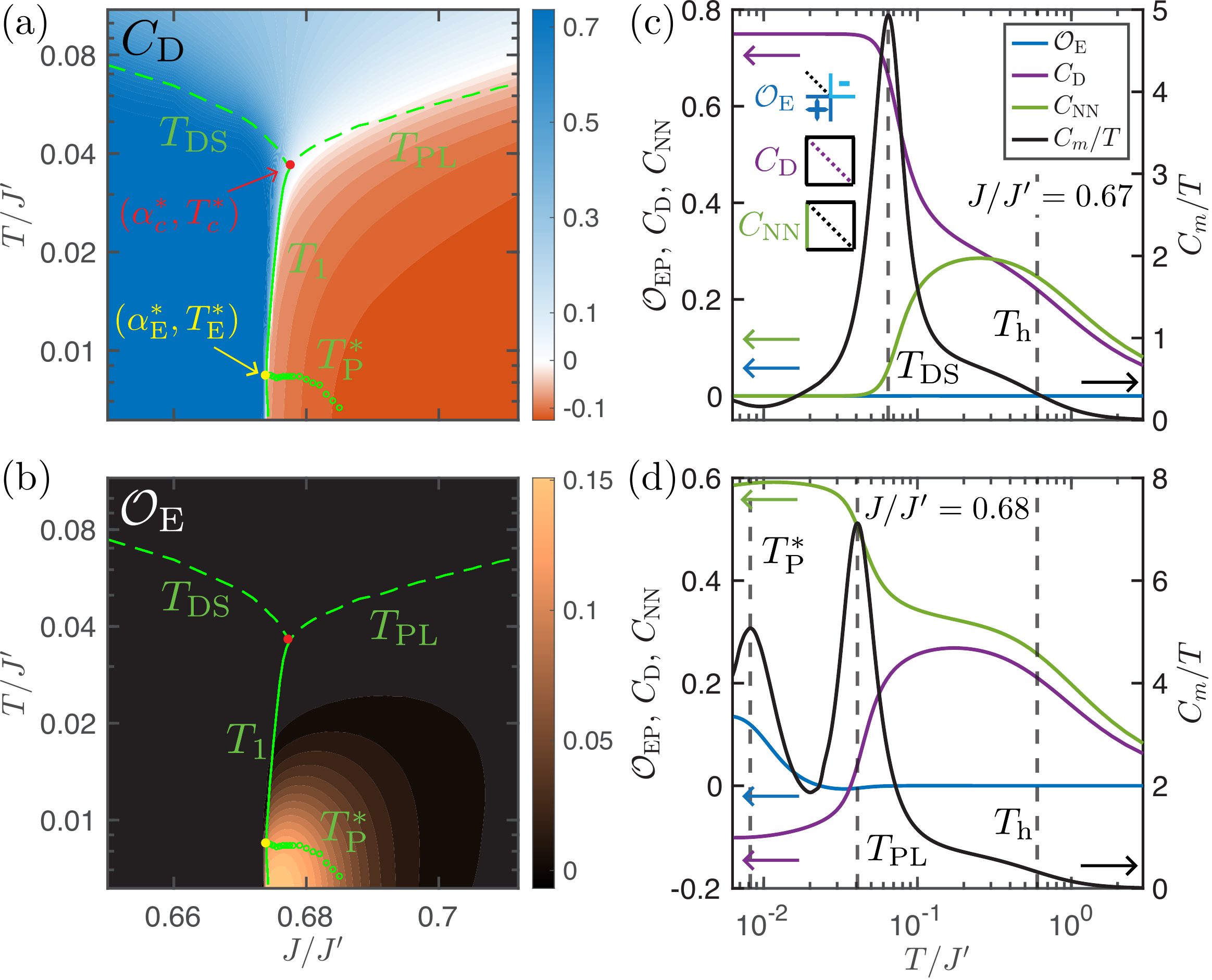}
\caption{Contour plots of (a) intra-dimer correlator $C_{\rm D}$ and
(b) order parameter $\mathcal O_{\rm E}$ for the empty PS order. Green dashed, solid, 
and dotted lines are peaks of $C_m/T$, and red (yellow) dot represents 
the CP (CEP) [c.f. Fig.~\ref{fig1}(a)]. (c)-(d) Magnetic specific heat $C_m/T$, 
in conjunction with $C_{\rm D}$, inter-dimer correlator $C_{\rm NN}$, 
and $\mathcal O_{\rm E}$ v.s. $T$, for $J/J' = 0.67$ and $0.68$, 
respectively. The inset in (c) illustrates definitions of these quantities. 
Roughly $6\%$ total entropy is released near the $T_{\rm P}^*$ peak, 
consistent with experimental result of $\sim 4\%$ obtained at 
$1.8$~GPa~\cite{guo2020}.
}
\label{fig2}
\end{figure}

\textit{The SS model phase diagram.---}
The obtained pressure-temperature phase diagram of the SS model is 
shown in Fig.~\ref{fig1}(a) based on the contour plot of the magnetic specific 
heat $C_m/T$. As the first-order transition line is slightly bent, one goes 
over various spin states as temperature decreases with fixed $J/J'= 0.676$,
and find in Fig.~\ref{fig1}(c) intriguing temperature-evolution behaviors:
Starting from the high-$T$ paramagnetic (PM) phase, the system evolves
into the DS regime [second row of Fig.~\ref{fig1}(c)], where the intra-dimer
correlation $C_{\rm D}=-\expval{\vb S_i\cdot \vb S_j}_{\rm D}$ [c.f., inset 
in Fig.~\ref{fig2}(c)] is strongest and the spin structure factor peaks at $X$ 
and $Y$ points in the first Brillouin zone. Further decreasing temperature, 
it enters the plaquette liquid (PL) phase via a first-order transition, where 
$C_{\rm D}$ changes its sign and the inter-dimer correlation $C_{\rm NN}
=-\expval{\vb S_i\cdot \vb S_j}_{\rm NN}$ [see also inset in Fig.~\ref{fig2}(c)] 
becomes stronger, as shown in the third row of Fig.~\ref{fig1}(c). In the PM, 
DS, and PL regimes, the equivalent NN bonds take the same values and 
there is no $\mathbb{Z}_2$ symmetry breaking; while at sufficiently low 
temperatures ($T < T_{\rm P}^*$), the PL phase eventually gives way to 
the ordered PS phase upon a second-order transition. The PS order can 
be detected by comparing two bonds with the same orientation
\footnote{The cylinder boundary condition introduces an effective pinning 
field that may favour a particular plaquette pattern~\cite{yang2022}, and 
in practice we measure the order parameters in the central unit cell as 
elaborated in~\cite{SM}.}, as shown in the last row of Fig.~\ref{fig1}(c).
Correspondingly, the spin structure peak shifts to the $M$ point in the PL
and ordered PS phases, with the latter being brighter. Markedly, this 
interesting temperature evolution of spin states due to the slightly bent
first-order line is consistent with the results in a recent NMR experiment, 
where a phase coexistence phenomenon was observed~\cite{cui2022}.

\emph{First-order line and critical point.---}
We determine the critical point (CP), $(\alpha^*_{\rm c}, T^*_{c}/J') \simeq 
(0.678, 0.036)$, in Fig.~\ref{fig1}(a) as the position where two dashed lines, 
$T_{\rm DS}$ and $T_{\rm PL}$, 
and the solid line $T_{1}$ meet. The latter, i.e., a first-order transition line, 
can be understood from intra-dimer correlation $C_{\rm D}$, which serves 
as the corresponding density-type order parameter. As shown in Fig.~\ref{fig2}(a), 
for $T<T^*_c$, a discontinuous jump in $C_{\rm D}$ occurs at $\alpha_1 \simeq
0.675$, a characteristic of the first-order transition between the DS and PL/PS 
phase. In contrast, for $T>T^*_c$, the simulated $C_{\rm D}$ data show a 
smooth change, which resembles that of liquid-gas crossover in the supercritical 
regime of water's pressure-temperature phase diagram~\cite{jimnez2021}. 
According to the couplings determined in Refs.~\cite{zayed2017,guo2020}, 
our results correspond to a critical pressure around 2~GPa, and {a critical 
temperature} $T_c^*$ about 2-3~K, in agreements with recent experiments
\cite{jimnez2021}. We further examine the correlation jump between the DS 
and PS states $\Delta C_{\rm D}\sim (\frac{T_c^\ast-T}{T_c^\ast})^\beta$ near 
the CP (from below)~\cite{jimnez2021}, with the fitted critical exponent 
$\beta\approx1/8$~\cite{SM} falling into the two-dimensional Ising universality 
class. Notably, the white line in Fig.~\ref{fig2}(a) with $C_{\rm D}\approx 0$, 
which roughly coincides with the $T_{\rm PL}$ line determined independently 
from the broad peak of $C_m/T$, indicates a \emph{sign switching} in the 
intra-dimer correlations, as also observed experimentally~\cite{bettler2020}.

In Fig.~\ref{fig1}(a), there are two crossover temperature scales, $T_{\rm DS}$
and $T_{\rm PL}$, determined from $C_m/T$ humps. They can be elucidated 
by examining two types of local correlations, i.e., the inter-dimer $C_{\rm NN}$
and the intra-dimer $C_{\rm D}$. As shown in Fig.~\ref{fig2}(c) and (d), 
a shoulder-like structure is firstly developed  in $C_m/T$ at $T_{\rm h} \sim 0.4 J'$, 
where both correlations build up with similar strengths. However, it is found that 
for $\alpha =0.67<\alpha_1$, [c.f., Fig.~\ref{fig2}(c)], the intra-(inter-) dimer 
correlations increase (decrease) rapidly around $T_{\rm DS}$; while, for 
$\alpha = 0.68>\alpha_1$, [c.f., Fig.~\ref{fig2}(d)], the situation around 
$T_{\rm PL}$ is reversed. Moreover, we define the empty PS order 
parameter $\mathcal O_{\rm E}\equiv \sum_i \bqty{(-1)^{i_x} \vb S_i \cdot \vb 
S_{i+\hat x} - (-1)^{i_y} \vb S_i \cdot \vb S_{i+\hat y}}$, with $i_{x,y}$ the 
coordinates of site $i$ [c.f., Fig.~\ref{fig1}(b)], and the summation runs 
over sites of the central unit cell to alleviate finite-size effect. It is found that 
$\mathcal O_{\rm E}$ remains vanishingly small till near $T_{\rm P}^\ast$, 
indicating the fluctuating plaquette order in the PL regime yet without 
$\ZZ$ symmetry breaking.

\emph{Second-order line and critical endpoint.---}
When further decreasing temperature, the specific heat peak gets brighter
and becomes maximal at $(\alpha_{\rm E},T_{\rm E}/J') \approx (0.674, 0.008)$,
as shown in Fig.~\ref{fig1}(a), which is nothing but the critical endpoint (CEP)
\cite{fisher1990,fisher1991}. On the right hand side of this point, there is a
second-order thermal transition line $T^\ast_{\rm P}$ defined by the peaks
of $C_m/T$, which goes downwards and eventually drops outside of our
temperature window when approaching the possible QPT at $\alpha_2^\ast
\approx 0.692$ determined by DMRG calculations~\cite{SM}. Such a peak
structure has been experimentally observed recently~\cite{guo2020,jimnez2021}. 
Here we perform low-temperature calculations down to previously inaccessible 
regime~\cite{wessel2018,wietek2019,jimnez2021,shimokawa2021}, and identify 
the ordered PS phase below the transition temperature $T_{P}^* \sim 0.01 J'$,
i.e.,  resembling the ``solid'' or ``ice'' beneath the ``liquid'' phase in the water's 
phase diagram. The low-temperature calculations smoothly extrapolate to the 
ground-state DMRG results, and hence provide a \emph{comprehensive} 
pressure-temperature phase diagram of the SS model in Fig.~\ref{fig1}(a). 
Noteworthily, although the PS order transition found experimentally is 
around $T/J' \sim 0.02$~\cite{guo2020,jimnez2021}, slightly higher than the 
value obtained here for $W=6$ geometry, our finite-size analysis indicates 
that $T_{\rm P}^\ast$ increases with width, and the rudimentary width-8 result 
of  $T_{P}^*$ already takes a similar value (c.f., Supplemental Fig.~S2~\cite{SM}).

To understand the nature of this low-temperature PS phase, in particular, 
whether the symmetry breaking occurs among the empty or full plaquettes,
we compute order parameters for both the empty ($\mathcal O_{\rm E}$)
and full ($\mathcal O_{\rm F}$) PS states. The former is shown in Figs.~\ref{fig2}(b-d), while the latter is found to be much smaller (see more details
in Supplemental Sec.~III Subsec.~B~\cite{SM}). 
In Fig.~\ref{fig2}(d), we find $\mathcal O_{\rm E}$ remains zero until 
around $T^\ast_{\rm P}$, where the specific heat shows a peak. 
It corresponds to the rapid buildup of $\ZZ$ symmetry breaking order 
amongst empty plaquettes. We also compute the PS susceptibility 
for both empty {($\chi_{\rm E}$)} and full {($\chi_{\rm F}$)} PS orders, 
and find that the former increases much faster even in the PL regime
below $T_{\rm PL}$~\cite{SM}. Therefore, we conclude that the empty PS
instability predominates over the full one in the entire low-temperature range, 
and confirm that the PS phase is of empty-type in the pressure-temperature
phase diagram. Since recent experiments indicate instead the 
full-plaquette state in {\scbo}~\cite{zayed2017,bettler2020,cui2022}, 
one needs to consider additional terms beyond the basic SSM to 
resolve this subtle discrepancy~\cite{boos2019}.

\begin{figure}[!t]
\centering
\includegraphics[width=\linewidth]{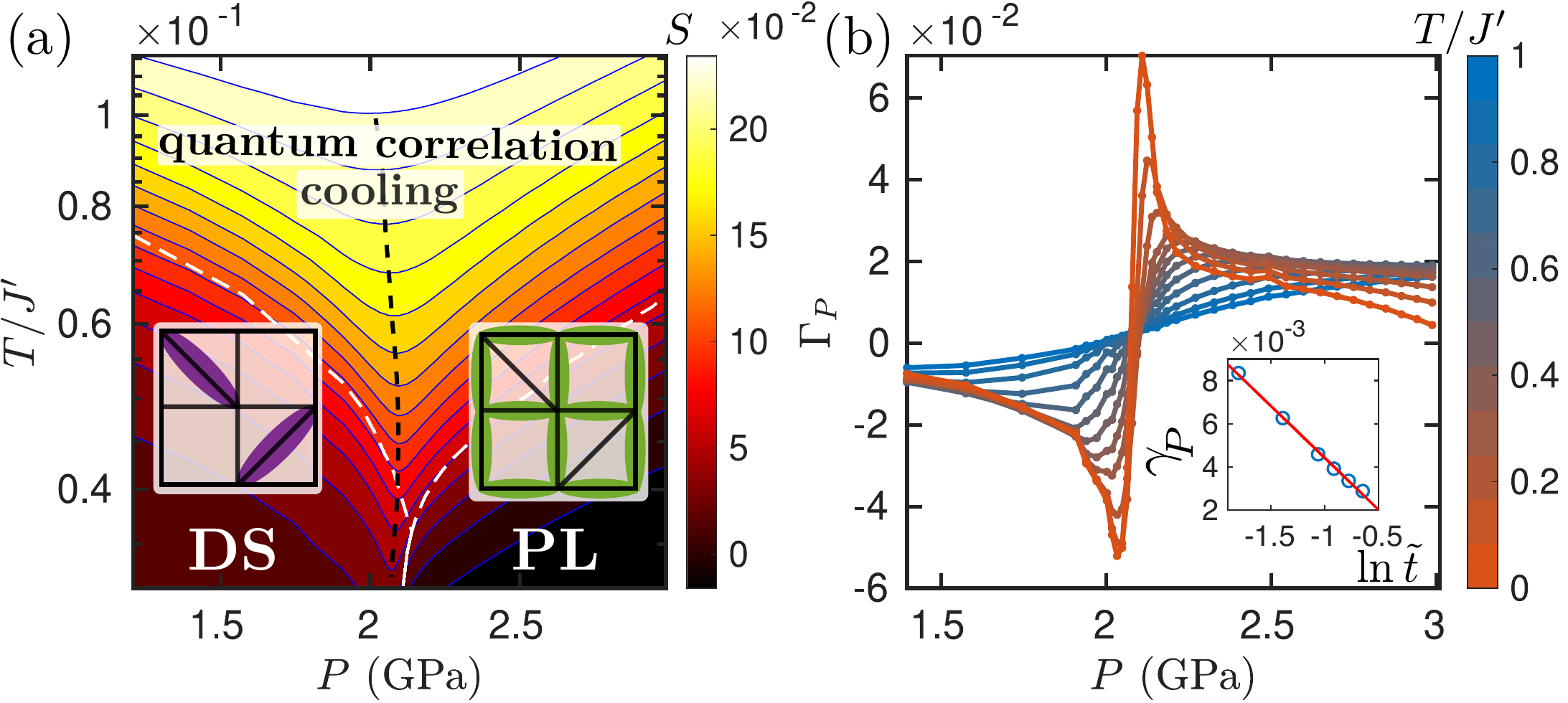}
\caption{(a) Contour plot of thermal entropy $S$ with isentropes
indicated by blue solid lines, {where prominent cooling effect
due to dramatic change in quantum correlations is observed.}
The white dashed and solid lines are determined from $C_m/T$
[c.f., Fig.~\ref{fig1}(a)]. The black dashed line connects the dips
of isentropes and represents a line of maximal entropy.
(b) {\gr} $\Gamma_P$ vs. pressure $P$. The inset shows 
$\gamma_P(\tilde{t})$ v.s. $\ln{\tilde{t}} \equiv \ln{[(T-T_c^\ast)/T_c^\ast]}$ 
(blue circles), with the fitting (red line) also shown. In both panels, 
we use the relation between coupling strength and pressure given 
by~\cite{zayed2017}, i.e., $J'(P) = (75-8.3P/\rm{GPa}) \rm{K}$ and 
$J(P)=(46.7-3.7P/\rm{GPa}) \rm{K}$.}
\label{fig3}
\end{figure}

\emph{Supercritical regime and magnetic barocalorics.---}
It is well-known that water in its supercritical state has many fascinating
physical properties leading to various applications~\cite{clifford2000}.
In the case of quantum magnets, the supercritical regime remains largely 
unexplored~\cite{weber2022}. Here we initiate the investigation of 
supercriticality in the SS model from the perspective of magnetothermodynamics. 
In Fig.~\ref{fig3}(a), we present the isentropes in this regime, where 
a prominent adiabatic magnetic cooling effect is found. By connecting 
the lowest temperature points of isentropes, we obtain a maximal entropy 
line with strong spin fluctuations that resembles the renowned Widom 
line in the supercritical regime~\cite{luo2014}. As such a cooling effect 
originates from the magnetic entropy change and is controlled by pressure, 
we dub it \emph{magnetic barocaloric effect} (mBCE), and propose 
to characterize it by a {\gr} $\Gamma_P \equiv-\frac{1}{T} 
\frac{(\partial S/\partial P)_T}{(\partial S/\partial T)_P}$. As shown 
in Fig.~\ref{fig3}(b), a clear sign change with very pronounced
peak/dip can be observed in $\Gamma_P$ for supercritical spin states.
More specifically, we denote the nominator by $\gamma_P \equiv
-(\partial S/\partial P)_T \sim \partial\expval{\vb S_i\cdot \vb S_j}/\partial T$,
and find a universal scaling $\gamma_P \sim \ln{[(T-T^*_c)/T^*_c]}$
near the CP  [c.f., the inset of Fig.~\ref{fig3}(b)], dictated by the 2D Ising 
universality class. Notably, in sharp distinction to the conventional magnetic 
cooling due to entropy change via order-disorder
switch of magnetic moment's orientations~\cite{deoliveira2014}, 
here the mBCE is related to the {rearrangements} in spin singlet 
patterns [illustrated in the inset of Fig.~\ref{fig3}(a)]. Such a quantum 
correlation induced cooling, observed in the supercritical regime and 
ascribed to the rearrangement of disorder singlet patterns, constitutes a novel mechanism
for helium-free cryogenics.

\begin{figure}[!t]
\centering
\includegraphics[width=1\linewidth]{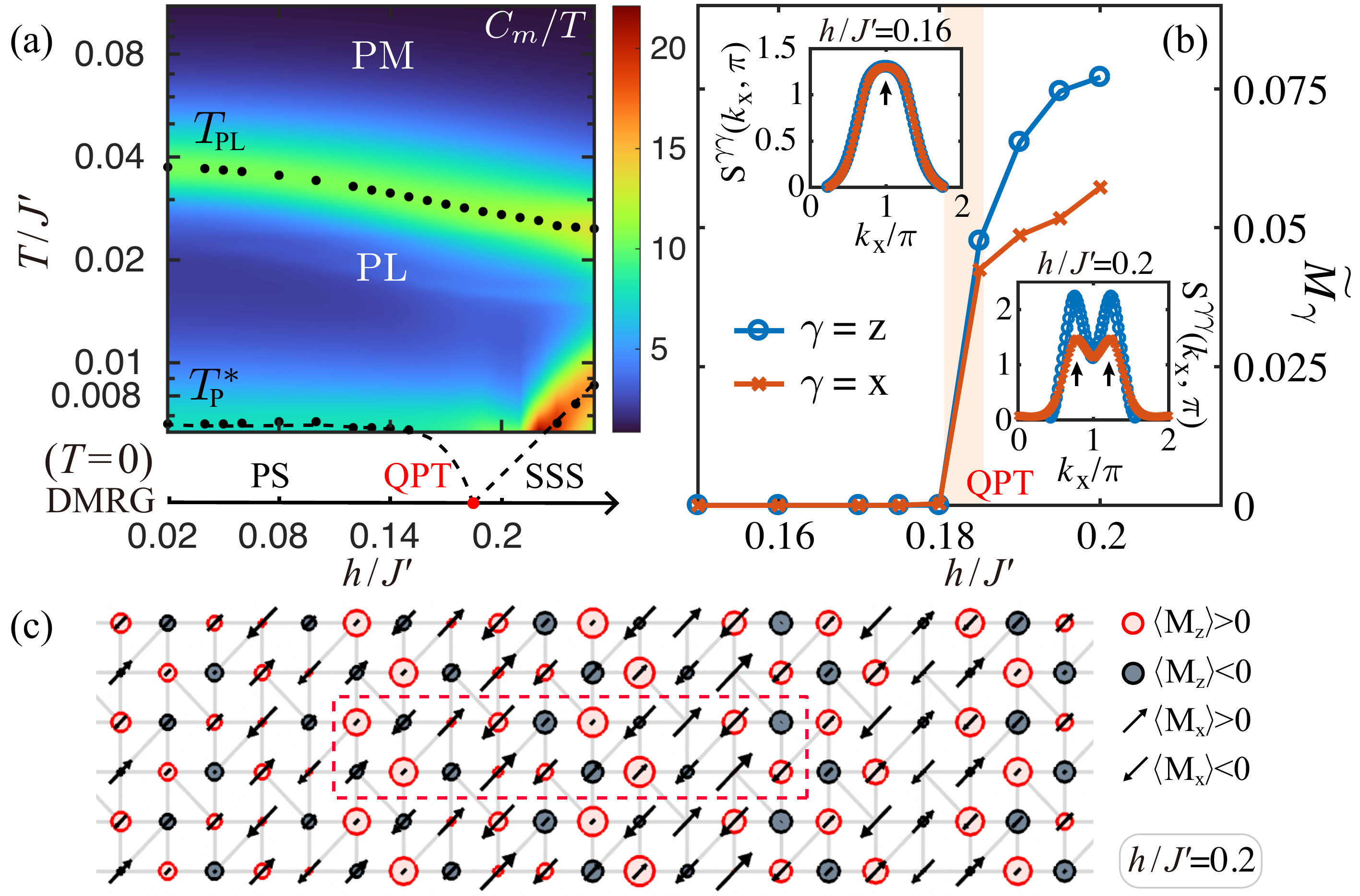}
\caption{(a) Contour plot of $C_m/T$ with black dots indicating 
locations of the peaks, which extrapolate to the ground-state 
phase diagram where a QPT occurs at $h_{c}/J' \simeq 0.185$
(for $J/J'=0.68$). 
(b) The ``solid'' and ``superfluid'' spin order parameters 
$\tilde{{M}}_{\gamma}$ ($\gamma = x,z$) computed as 
$\frac{1}{N_{\rm Bulk}}\sum_{i\in {\rm Bulk}} |\langle M_{\gamma} 
\rangle_i|$ with site $i$ running over bulk of the cylinder.
The insets show the spin-structure factors $S^{zz}(k_x,k_y)$ and
$S^{xx}(k_x,k_y)$, under two different fields $h/J'=0.16$ and $0.2$.
(c) Spin texture in the SSS phase, where the moments are displayed
with the longitudinal ($\langle M_z \rangle_i$) and transverse
($\langle M_x \rangle_i$) magnetic moments. The positive (negative)
$\langle M_z \rangle_i$ values are illustrated with red (black) circles,
and the $\langle M_{x} \rangle_i$ components with the arrows. 
Size of circles and length of arrows reflect the absolute values 
of corresponding local moments, where a $10\times2$ unit cell
on a 6$\times$30 cylinder is indicated by the red dashed box.
}
\label{fig4}
\end{figure}

\emph{{Field-temperature phase diagram and} spin-supersolid transition.---}
Given the ordered PS phase identified, we consider {applying magnetic
fields to pressured} \scbo~\cite{shi2022,cui2022} along the spin $S^z$
direction, i.e., $H \rightarrow H-h\sum_i S_i^z$. Here we focus on $J/J'=0.68$ 
and $h/J'\leq0.25$, and the contour plot of $C_m/T$ is shown in Fig.~\ref{fig4}(a), 
where we find both temperature scales, $T_{\rm PL}$ and $T^\ast_{\rm P}$, 
decrease as the field increases, and the latter will eventually drop out of the 
available temperature window. A low-temperature $C_m/T$ peak reappears for $h/J' 
\gtrsim 0.22$, suggesting {a QPT occurs before $h/J'=0.22$}.

To clarify {the quantum phases and phase transitions} in Fig.~\ref{fig4}(a),
we perform DMRG calculations and show the results in Figs.~\ref{fig4}(b) 
and (c). On width-6 cylinder, there exists a pressure-induced intermediate 
PS phase~\cite{lee2019}, which gives way to a stripy SSS~\cite{shi2022} 
phase for $h > h_c/J'\simeq 0.185$ [c.f., Fig.~\ref{fig4}(b)] via a QPT 
possibly of first order~\cite{SM}. In the SSS phase with $h/J' = 0.2$, 
we show the computed local moments in {Fig.~\ref{fig4}(c)} where a 
$10\times2$ unit cell can be observed, similar to the previously iPEPS 
results~\cite{shi2022}. In the SSS phase, both $\langle M_x \rangle$ and 
$\langle M_z \rangle$ are nonzero, indicating that both translational and 
U(1) symmetries are simultaneously broken, i.e., there exists a quantum 
magnetic analogue of supersolidity. Taking $J' \approx 60$~K for \scbo~under
pressure of about 2.0~GPa~\cite{guo2020}, we estimate the field-driven 
QPT takes place at $h_c\approx 8$~T, in agreement with recent 
experiments~\cite{shi2022,cui2022}.

The SSS order is also evident in the spin structure factors $S^{\gamma
\gamma}(\vb k) = N^{-1}\sum_{i,j} e^{-i\vb k \cdot (\vb r_i-\vb r_j)} \langle
S_i^\gamma \cdot S_j^\gamma \rangle$ (with $N$ the total lattice sites)
shown in the insets of Fig.~\ref{fig4}(b), where a broad peak in the PS 
phase changes into a double-peak structure split 
apart at $k_x = \pi \pm \frac{\pi}{5}$, as $h$ changes from $0.16$ 
to $0.2$. Such a peak-splitting behavior is also found when 
decreasing temperature~\cite{SM}, accessible by neutron scattering 
measurements for probing the SSS phase. We note that for
different choices of $\alpha$, other phases may show up instead 
of SSS, making the field-driven spin states and transitions
extremely rich in the pressured \scbo~\cite{shi2022,cui2022}.

\emph{Discussion and outlook.---}
Recent experimental advances~\cite{guo2020,jimnez2021,cui2022}
have added greatly to the understanding of the pressure-field-temperature 
phase diagram of {\scbo}, which advocate comprehensive theoretical 
studies. Here with the state-of-the-art tensor-network 
approach, we map out the finite-temperature phase diagram that explains 
experimental findings and opens a refreshing avenue in the {\scbo} studies. 
In particular, the ordered PS phase and its thermal transition line are 
identified, and the nature of the \emph{low-temperature} PS phase
is clarified by finding that the empty PS order is always predominant 
over the full one at low temperature. This nails down the direction 
to explain the discrepancy, i.e., the original SS model could not fully capture 
the low-$T$ phase in \scbo, and one might need to consider additional terms 
for the model Hamiltonian. As the PS transition takes place at very low 
temperature, i.e., $T^*_{\rm P}/J' \sim O(0.01)$, it suggests that other 
(small) interactions, including the interlayer 
couplings ($\lesssim 10\%$ of $J'$~\cite{miyahara2003}), spin-orbit couplings 
($\sim 3\%$~\cite{nojiri2003}), distortion~\cite{boos2019}, and staggered 
ring exchange interaction~\cite{xi2023}, may be relevant to make more 
pertinent theoretical explanation on {\scbo}.  Moreover, we call for future 
experimental investigations of the magnetic barocalorics in the supercritical 
regime, which constitutes a \textit{bona fide} spin correlation cooling effect 
fundamentally different from traditional magnetic refrigeration. The novel
cooling mechanism in frustrated magnets enables potential applications 
in space cryogenics~\cite{Shirron2014} and quantum technologies
\cite{Jahromi2019nasa}.

The pressure-driven PS-AF quantum phase transition has been 
intensively studied recently~\cite{lee2019,yang2022, wang2022}. 
Our finite-$T$ studies here, for width-6 system,
 find no salient feature near $\alpha_2^*$ in simulated 
quantities~\cite{SM}. Whether this transition belongs to a 
DQCP~\cite{lee2019} or is replaced by a QSL phase~\cite{yang2022,
keles2022,wang2022} unfortunately cannot be addressed here. Instead, 
the field-driven PS-SSS transition in the case of $\alpha=0.68$ (i.e., 
2.2~GPa pressure in experiments) belongs to first-order and can 
be probed by magnetocaloric measurements~\cite{SM}. Meanwhile, 
a field-driven PS-AF phase transition was recently observed under a 
relatively higher pressure (e.g., $2.4$ GPa) and has been suggested 
to be a proximate DQCP~\cite{cui2022}, which remains for future studies. \nocite{li2023,gao2023,miles2012,weichselbaum2012,weichselbaum2020,
dong2017,zhu2003,garst2005,haegeman2016,haegeman2011}

\begin{acknowledgments}
\textit{Acknowledgments.---}
J.W. and W.L. are indebted to Rong Yu, Ling Wang, Anders Sandvik, Zi 
Yang Meng, and Weiqiang Yu for stimulating discussions. This work was 
supported by the National Natural Science Foundation of China (Grant 
Nos.~12222412, 11834014, 11974036, and 12047503), National Key 
R\&D Program of China (Grant Nos.~2018YFA0305800 and 
2022YFA1402704), Strategic Priority Research Program of Chinese 
Academy of Sciences (CAS) (Grant No.~XDB 28000000), the 
Fundamental Research Funds for the Central Universities, the CAS
Project for Young Scientists in Basic Research (YSBR-003,YSBR-057), and the Innovation Program for Quantum Science and Technology (under Grant No. 2021ZD0301900). 
We thank the HPC-ITP for the technical support and generous allocation of CPU time.
\end{acknowledgments}

%

%
\newpage
\clearpage
\onecolumngrid
\mbox{}
\begin{center}
\textbf{\large Supplemental Material}
\end{center}

\date{\today}

\setcounter{section}{0}
\setcounter{figure}{0}
\setcounter{equation}{0}
\renewcommand{\theequation}{S\arabic{equation}}
\renewcommand{\thefigure}{S\arabic{figure}}
\setcounter{secnumdepth}{3}


\section{Finite-temperature Tensor Renormalization Group}

\subsection{Convergence check of results}

\begin{figure}[b!]
\includegraphics*[width=.8\linewidth]{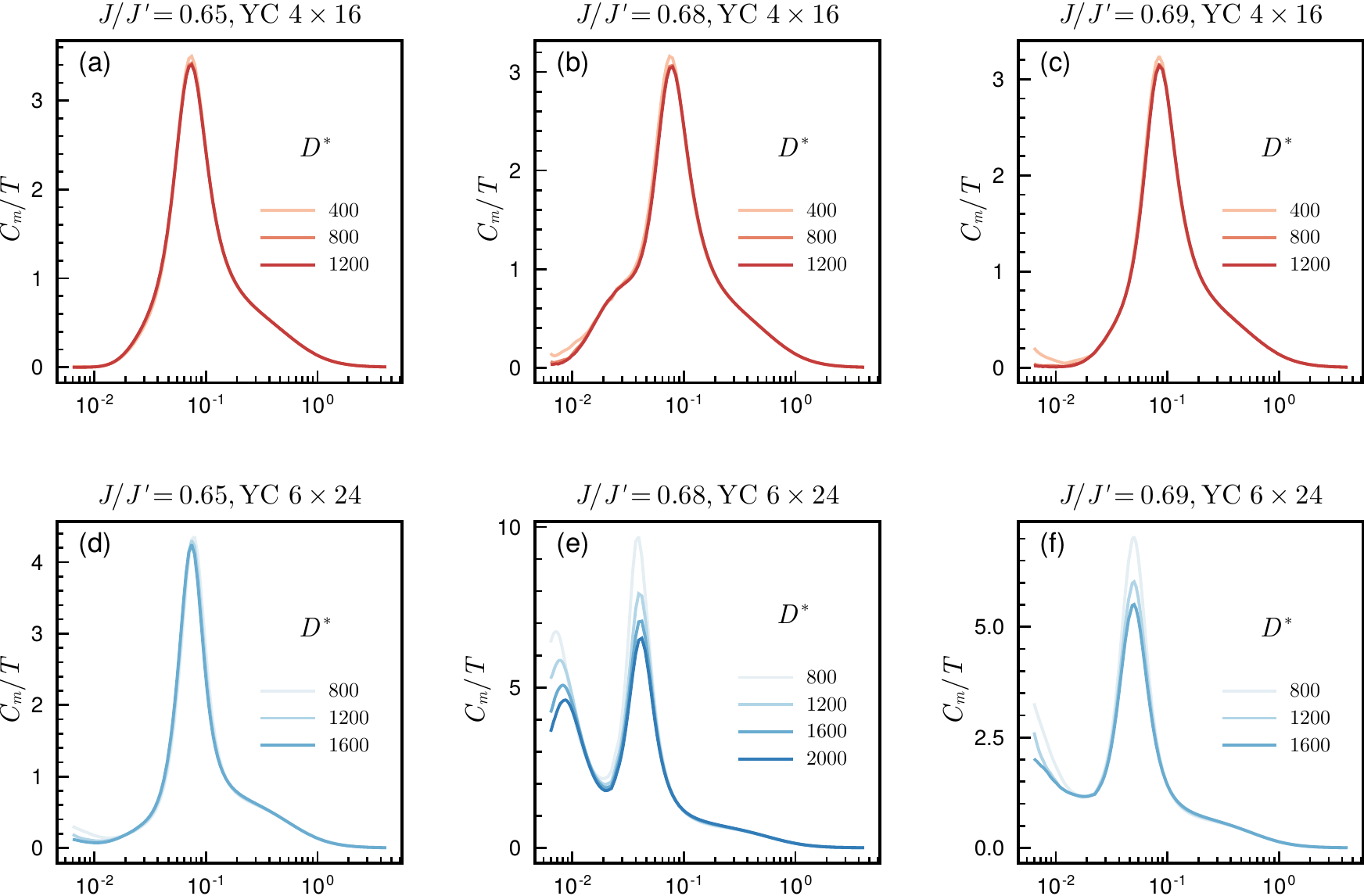}
\caption{Convergence of the tensor network simulations for the SS model 
on width-4 and 6 cylinders: Magnetic specific heat $C_m/T$, for system size $W \times L = 4\times 16$ 
(a-c) and $6\times 24$ (d-f).}
\label{smFig0}
\end{figure}

Here we present more details of the thermal tensor network approaches employed
in the present study of Shastry-Sutherland (SS) model, i.e., the exponential tensor 
renormalization group (XTRG) based on matrix product operators (MPOs). XTRG is
a highly controllable method that can reach low temperature exponentially fast. 
Starting with a quasi-1D mapping of the
original square lattice into a snake path, we represent the Hamiltonian $H$ in the
form of an MPO. At a very high temperature $\beta_0 \equiv 1/T_0 \ll 1$, the density
matrix $\rho_0(\beta_0)$ of the system can be expanded via series-expansion thermal tensor network (SETTN) approach~\cite{Chen2017} as,
\begin{equation}
  \rho_0(\beta_0) \simeq \sum_{n=0}^{n_c} \frac{(-1)^n \beta_0^n}{n!}H^n,
\end{equation}
with $n_c$ a small integer (less than 10 in practice).
For example, when $\beta_0 =10^{-6}$, $n_c=3$
suffices to result in very accurate representation essentially free of
expansion error. After that, we keep squaring the MPO~\cite{Chen2018}
\begin{equation}
  \rho_m(2^m \beta_0) = \rho_{m-1}(2^{m-1} \beta_0) \cdot \rho_{m-1}(2^{m-1} \beta_0),
  \label{eq:mpop}
\end{equation}
and the system cools down exponentially fast. As the MPO product in
Eq.~\eqref{eq:mpop}  leads to enlarged bond dimension, one actually needs to
compress the bond space. Naively, the numerical cost for the MPO compression
scales like $\mathcal O(D^6)$. However, a $\mathcal O(D^4)$ complexity can
be achieved with a variational optimization~\cite{Chen2017}. 

Very recently, a method {\ttrg} \cite{li2023} with $\mathcal O(D^3)$
complexity has been developed to tackle the challenging
problem like 2D Hubbard model. Here we exploit it to boost the calculations
of the SS model by retaining even larger bond dimensions for width-8 cylinders. 
The {\ttrg} method is an 
integrated framework based on the matrix product operator (MPO) representations: 
One first employs the SETTN technique~\cite{Chen2017}
to prepare an accurate initial density matrix (in MPO) at high temperature, 
designs an evolution procedure with bilayer construction~\cite{dong2017}, 
and then deploys the tangent-space technique, i.e., time-dependent variational principle (TDVP)~\cite{haegeman2011,haegeman2016}, to conduct the imaginary-time evolution. 
When the prescribed temperature is reached, one can compute the thermodynamics, 
e.g., specific heat and magnetic susceptibility, as well as various spin correlations, 
with high accuracy. 
The {\ttrg} method incorporates several techniques and enables an accurate and efficient 
calculation of finite-temperature properties of large-scale quantum many-body systems. 
More details and elaborative examples on this method applied to quantum spin systems 
will be described in a separate work~\cite{gao2023}.


As illustrated in Fig.~1(a), the SS model is put on a $W\times L$ cylindrical
lattice with $W$ being the width and $L$ the length. To relieve severe boundary
effects, instead of the standard $L/W=2$ aspect ratio~\cite{miles2012,yang2022},
we stick to a larger one, $L/W=4$ with fixed width $W=6$, and the simulations
are performed down to a very low temperature of $T/J' \sim 0.006$.
In zero (nonzero) magnetic field, the global spin SU(2) [U(1)] symmetry of the
MPO is exploited based on the QSpace tensor library~\cite{weichselbaum2012,
weichselbaum2020}. The calculations are performed by retaining up to $
D^\ast = 2000$ multiplets, equivalent to $D\approx 6000$ U(1) states. In practice,
we always firstly ramp up $\beta$ exponentially, then followed by either an
exponential increase using XTRG, or a linear increase using {\ttrg}. Along
the way, utilizing a bilayer tensor trace~\cite{dong2017}, all thermodynamic
quantities can be obtained with high accuracy.

In Fig.~\ref{smFig0}, we show the magnetic
specific heat $C_m/T$ for several different bond dimensions $D^\ast$ and system width $W=4$ and 6. For the former case, it is
found that the convergence is well reached for $D^\ast=1200$; while for the latter case,
the convergence is pretty much reached for $D^\ast = 1600$ (or 2000 for
the more challenging case near the DS-PS transition), which is used in our
numerical results presented in the main text.

\subsection{Finite-size analysis of cylinder simulations with widths $W=4,6,8$}
In the main text, we only present results for cylinders with width $W=6$. Here 
we discuss results for cases of $W=4$ and $8$ to analyze the finite size effect in our calculation.

First, we notice that Ref.~\cite{lee2019} using iDMRG method has pointed out that $W=6$ 
is the \textit{minimal} width which hosts an intermediate plaquette singlet (PS) ordered phase in the
ground state, and this PS phase range increases with width, see Table~\ref{tab:table1}%
\begin{table*}
\caption{\label{tab:table1}Two zero-temperature transition points of the SSM in zero magnetic field, for the DS-PS and PS-AF transition, respectively. Data is obtained in Ref.~\cite{lee2019}.}
\begin{ruledtabular}
\begin{tabular}{c c c c c c}
&
$W=6$&
$W=8$&
$W=10$&
$W=12$\\
\colrule
$\alpha_1$ (DS-PS) & \text{$0.682$\footnote{Our finite-length width-6 result, $\alpha_1 = 0.673$, is more consistent with the thermodynamic result.}} & 0.677 & 0.675 & 0.675\\
$\alpha_2$ (PS-AF) & 0.693 & 0.728 & 0.762 & 0.77\\
\end{tabular}
\end{ruledtabular}
\end{table*}%
. Here we also examine the $W=4$ case, from the specific heat results [Fig.~\ref{smFig0}(b)], 
and conclude that there is no signal for the PS transition close to the expected transition point 
(despite a small hump near $T/J'\sim 0.025$), which is consistent with the ground-state results.

\begin{figure}
\includegraphics[width=\linewidth]{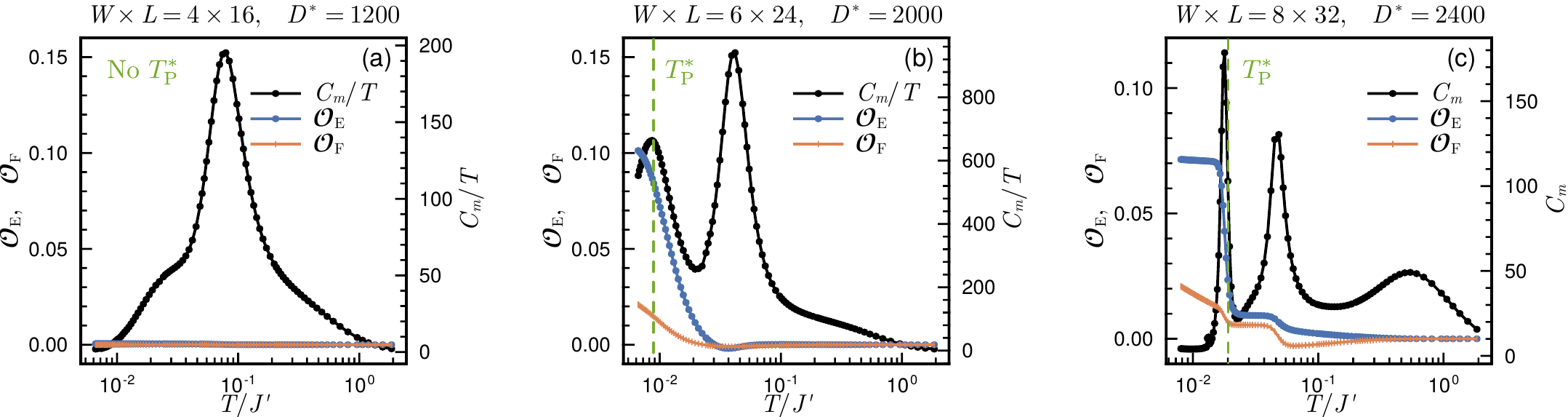}
\caption{The specific heat $C_m$ (black line {with dots}) and two types of PS order parameters, $\mathcal{O}_{\rm E}$ for empty plaquette ({blue} line {with dots}) and $\mathcal{O}_{\rm F}$ for full plaquette ({red} line {with vertical bar}), of the SSM in zero magnetic field for $J/J'=0.68$. The detailed definitions of these two order parameters are given in Sec.~\ref{defop}. Note for panel (c), the $W\times L = 8 \times 32$ case, 2400 multiplets, i.e., roughly $10000$ U(1) states, are kept in the simulation, while the full convergence of the data are not guaranteed here for $W=8$. Green dashed lines in (b) and (c) denote the temperature scale $T_{\mathrm P}^\ast$ corresponding to the PS transition, while there is no $T_{\mathrm P}^\ast$ for (a).
}
\label{yc8}
\end{figure}

We further compute one particular parameter, $J/J'=0.68$, for system size $W\times L = 8\times 32$ 
and bond dimension $D^\ast = 2400$. This computation is much more expensive, and is made 
possible by exploiting the algorithm {\ttrg} developed very recently \cite{li2023}. As can be seen from 
Fig.~\ref{yc8}(c), the black line for the specific heat $C_m$ shows two peaks at low $T$. It is found 
that only after the second peak, the empty PS order parameter ({blue} line) increases much faster than 
the full one ({red} line), meaning that the $\mathbb{Z}_2$ symmetry breaking occurs there. Note that 
both order parameters are not strictly vanishing even at higher temperatures, which may be ascribed 
to the anisotropy of cylinder geometry used, boundary effects, and also possibly the insufficiency of 
bond dimensions used in the very challenging width-8 simulations. Based on the rudimentary results 
in Fig.~\ref{yc8}(c), we estimate that the PS order transition temperature is {$T_{\rm P}^\ast \sim 0.02$} 
for $W=8$. Recalling that for $W=6$ and $J/J=0.68$, we have $T_{\rm P}^\ast / J' \simeq 0.01$ [c.f. 
Fig.~\ref{yc8}(b)] and there is no PS transition at all for $W=4$ [c.f. Fig.~\ref{yc8}(a)], it shows that 
\textit{the $T_{\rm P}^*$ becomes higher and the low-temperature PS phase enlarges with increasing 
widths}. In fact, two recent experiments find that the PS transition temperature $T_{\rm P}^\ast /J'$ 
locates around $0.02$~\cite{guo2020,jimnez2021}, close to the width-8 results here.


\section{Density matrix renormalization group simulations}
\label{smsec2}

In this section, we show the ground-state density matrix renormalization group
(DMRG) calculations of the Shastry-Sutherland (SS) model under both zero
and finite fields as a complement to the thermal tensor renormalization group
results.

\begin{figure}[h!]
\includegraphics[width=0.85\linewidth]{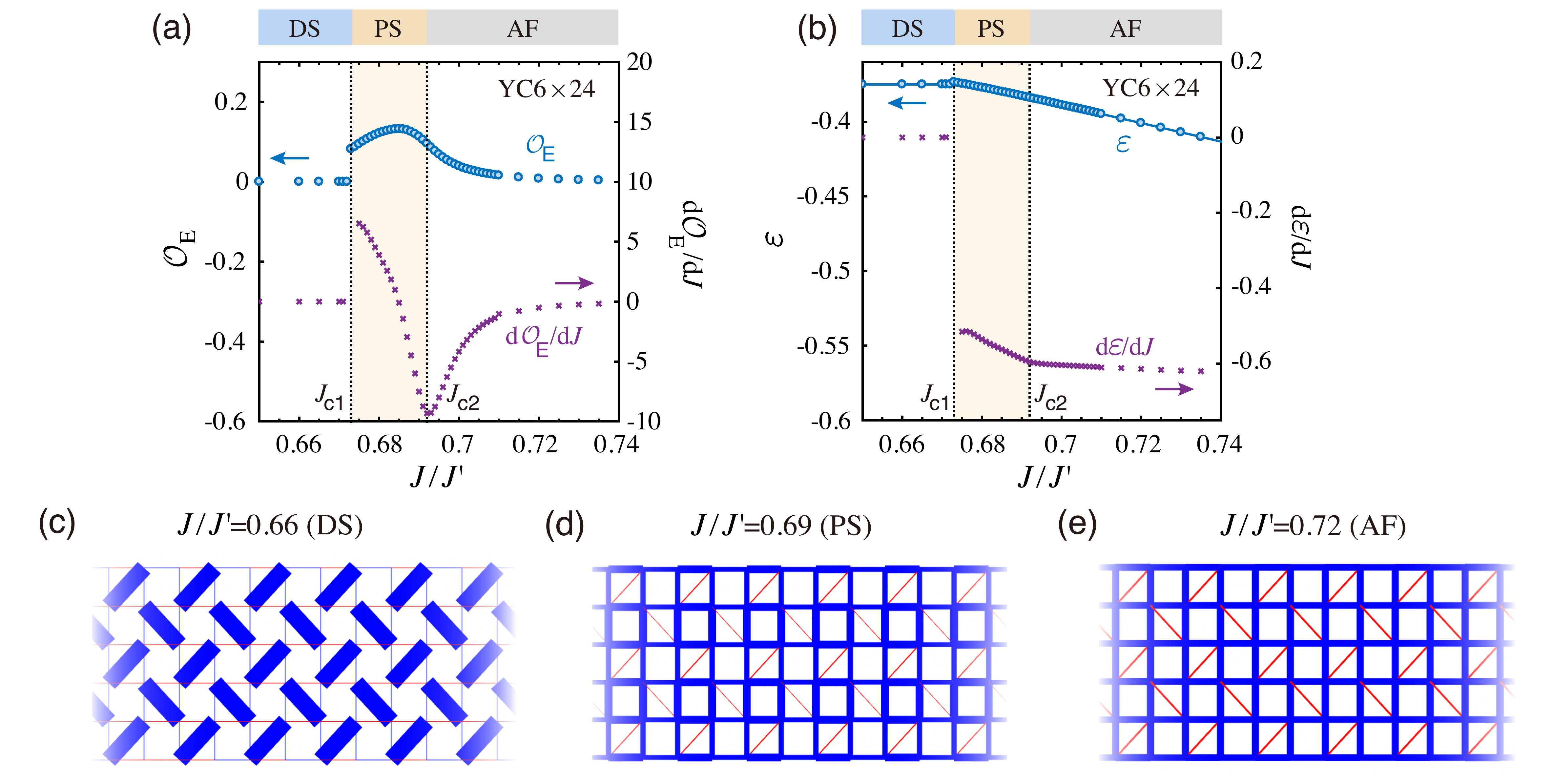}
\caption{Pressure-driven quantum phase transitions in the SS model
simulated on a $6\times24$ cylinder. (a) shows the PS order parameter 
$\mathcal O_{\rm E}$ and its derivatives ${\rm d} \mathcal O_{\rm E}/{\rm d}J$, 
where two QPTs separating the DS, PS, and AF phases can be observed 
at $J_{c1} \simeq 0.673$ and $J_{c2} \simeq 0.692$, respectively. 
(b) shows the ground-state energy per site $\varepsilon$ and its derivatives 
${\rm d}\varepsilon/{\rm d}J$ averaged over the central four sites of the lattice, where 
${\rm d}\varepsilon/{\rm d}J$ is found to be continuous and shows a kink at $J_{c2}$. 
(c-e) illustrate the three phases at representative coupling parameters, where 
the widths of the blue (red) bond indicate the negative (positive) values of the 
spin-spin correlations $\langle S_i S_j\rangle$.
}
\label{smFig:DMRG}
\end{figure}

\begin{figure}[h!]
  \includegraphics[width=0.75\linewidth]{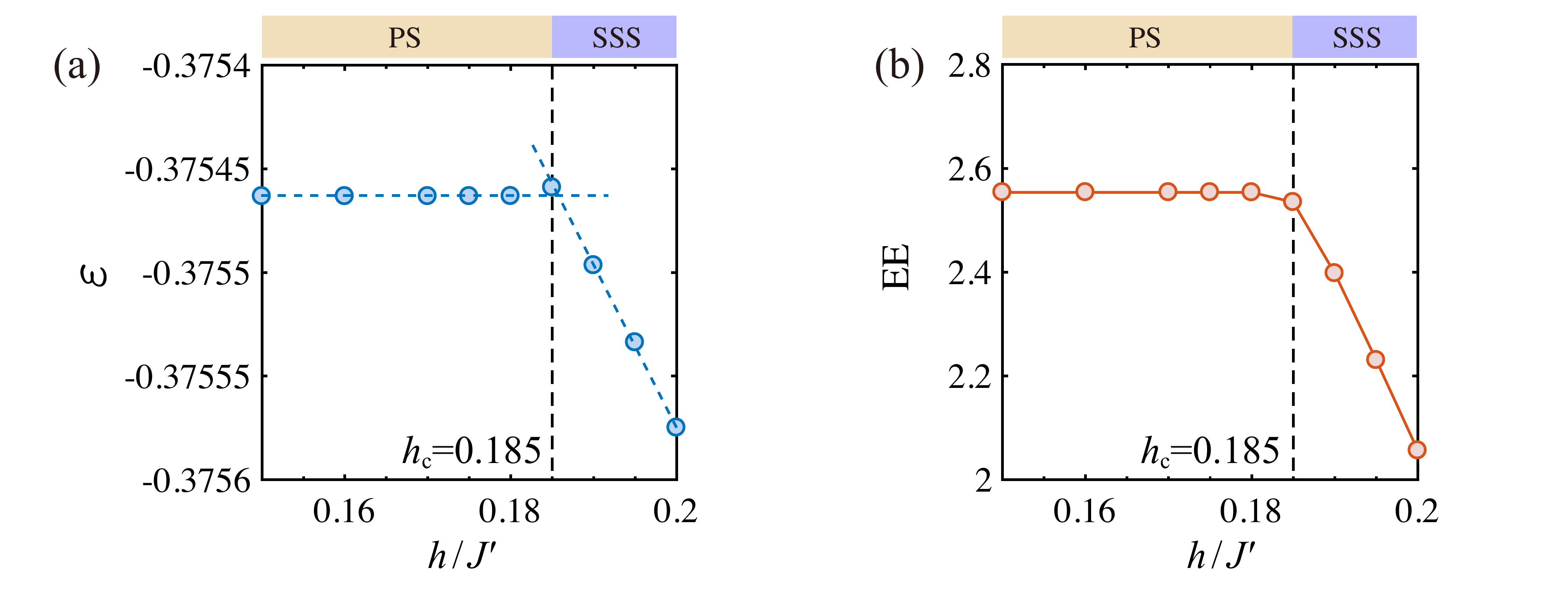}
  \caption{The ground state energy per site $\varepsilon$ and entanglement entropy (EE) 
  of SS model under magnetic fields. The calculations are performed on {a 
  $6\times 30$ cylinder} with $D = 1024$, where we fix $J/J'=0.68$. In (a), the 
  energy per site $\varepsilon$ shows a kink at $h_c\simeq 0.185$, and in (b) the EE starts 
  to decrease at $h_c\simeq 0.185$ and there is no divergence can be observed. 
  The results indicate that the QPT between the PS and SSS 
  phases is likely first order. 
  }
  \label{smFig:DMRGhz}
\end{figure}

\textit{Two quantum phase transitions of SS model at zero fields.---}
In Fig.~\ref{smFig:DMRG}, we calculate the empty plaquette singlet (PS)
order parameter $\mathcal O_{\rm E}$, the ground-state energy $\varepsilon \equiv
E/N_{\rm central}$ average over central sites of the lattice, and their
derivatives to determine the quantum phase transition (QPT), where the
simulations are performed on the $W\times L$ cylinder with width $W=6$
and length $L=24$. The kept bond dimension $D^*$ is up to 1024
multiplets, with SU(2) symmetry implemented~\cite{weichselbaum2012,
weichselbaum2020}, and the small truncation errors (below $\sim
10^{-7}$) guarantee high accuracies. During the calculations, the
intra-dimer Heisenberg coupling is fixed as $J' = 1$, and the NN
coupling $J$ is varied in the calculations, controlling the ratio $\alpha
\equiv J/J'$. As shown in Fig.~\ref{smFig:DMRG}(a), two QPTs from the
dimer singlet (DS) [illustrated in Fig.~\ref{smFig:DMRG}(c)] to the
PS phase [c.f., Fig.~\ref{smFig:DMRG}(d)] and then to antiferromagnetic
(AF) states [Fig.~\ref{smFig:DMRG}(e)] can be clearly identified,
which are labelled as $J_{c1}$ and $J_{c2}$, respectively. For small
$J$ values, the system is in the DS phase [c.f.~Fig.~\ref{smFig:DMRG}(c)],
and as $J$ is increased, the nonzero $\mathcal O_{\rm E}$ witnesses
the abrupt rise of the PS order. In Fig.~\ref{smFig:DMRG}(a), we find
$\mathcal O_{\rm E}$ shows a sudden jump at $J_{c1}\simeq 0.673$.
Such discontinuity is also seen for $\varepsilon$ as shown in
Fig.~\ref{smFig:DMRG}(b), indicating that the QPT at $J_{c1}$ is of first-order.

Further increasing $J$, the values of $\mathcal O_{\rm E}$ gradually
decrease as the system leaves the PS phase and enters the AF phase,
where a dip in ${\rm d} \mathcal O_{\rm E}/{\rm d}J$ appears at about
$J_{c2} \simeq 0.692$, i.e., the order parameter decreases most rapidly
at around  $J_{c2}$. This PS-AF QPT can also be seen in the results of
energy per site $\varepsilon$, whose first-order derivative is found to be continuous
in Fig.~\ref{smFig:DMRG}(b). There also exists a kink in ${\rm d}\varepsilon/{\rm d}J$
at $J_{c2} \simeq 0.692$, which also clearly locates the second-order
PS-AF transition.

\textit{The spin supersolid transition.---}
There are rich field-induced quantum spin states in SS model under magnetic
fields~\cite{shi2022}. Here in Fig.~\ref{smFig:DMRGhz}, for a typical $J = 0.68$,
we perform DMRG calculations on YC$6\times 30$ lattices and keep $D=1024$
states (with small truncation error $\sim 10^{-5}$) and find the field-induced spin
supersolid transition occurs at $h_c \simeq 0.185 J'$. At small magnetic fields
$h<h_c$, the system resides in the PS phase, where all the local moments
$\langle M_{\gamma} \rangle$ ($\gamma = x,y,z$) equal to zero, as discussed
in the main text. By increasing $h$, we find the system enters a spin supersolid
(SSS) phase {at $h_c$}. In Fig.~\ref{smFig:DMRGhz}(a) and (b), the energy per site $\varepsilon$ and
entanglement entropy (EE) curves are shown, respectively, where the kink in
energy and absence of divergence in EE results can be clearly observed at
$h_c$, indicating that the QPT is of first-order.

\section{More results on the SS model at zero magnetic field}

\subsection{Details on magnetic specific heat data}

\begin{figure}[h!]
\includegraphics*[width=0.7\linewidth]{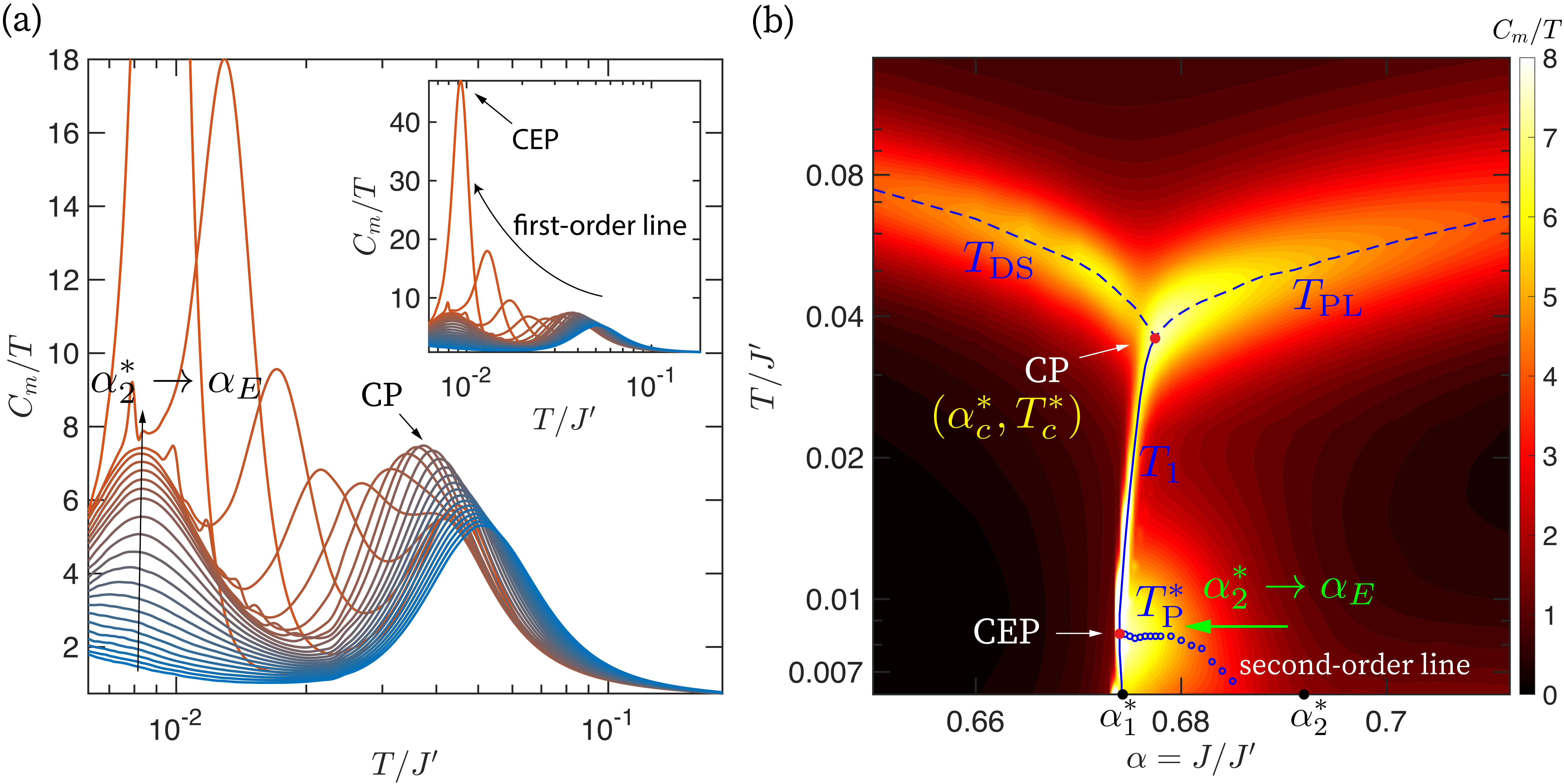}
\caption{(a) Magnetic specific heat $C_m/T$ v.s. temperature $T/J'$, for the coupling 
ratio $\alpha$ ranging from $\alpha_2^\ast \simeq 0.692$ to $\alpha_E=0.674$. 
(b) Contour plot of magnetic specific heat $C_m/T$.
The CP and CEP represent the critical point and critical endpoint of second-order phase
transitions, respectively. Note $\alpha_1^\ast$ and $\alpha_2^\ast$, as the ground-state DS-PS and PS-AF transition points, respectively, are determined from DMRG calculations.
}
\label{smFig:cv}
\end{figure}

Here we discuss in more details on magnetic specific heat data. In Fig.~\ref{smFig:cv}(a), we show $C_m/T$ v.s. $T/J'$, for the coupling ratio $\alpha$ ranging from $\alpha_2^\ast$ (PS-AF transition point) to $\alpha_E$ (critical endpoint). 
As $\alpha$ decreases, the high-$T$ hump (i.e., $T_{\rm PL}$) evolves into a peak labeled by the critical point (CP) around $T/J'\sim 0.04$ at $\alpha=\alpha_{\rm c}^\ast$. Further decreasing $\alpha$, this peak shifts to lower temperature (i.e., $T_1$) and becomes more pronounced. It eventually evolves into the critical endpoint (CEP) at $\alpha_{\rm E}$.
On the other hand, there is always a low-$T$ peak around $T/J' \sim 0.008$ (i.e., $T_{\rm P}^\ast$), which almost remains as a constant for $\alpha$ close to $\alpha_{\rm{E}}$. Moreover, for $\alpha<\alpha_{\rm c}^\ast$, an additional hump emerges at a temperature 
$T/J'$ higher than $0.04$ (i.e., $T_{\rm DS}$). We note that there are satellite subpeaks 
for $\alpha$ close to $\alpha_{\rm E}$, which are artifacts due to numerical errors, reflecting the fact that specific heat calculations are rather challenging at such low temperature. In Fig.~\ref{smFig:cv}(b), we show the contour plot of the magnetic specific heat $C_m/T$ in
a larger size and cleaner manner [as compared to Fig.~1(a) in the main text]. The blue 
circles at the $T_{\rm P}^*$ line, corresponding to peaks of $C_m/T$, are numerical 
data obtained in our calculation.

\subsection{Empty and full plaquette singlet order parameters}\label{defop}
Here we examine the full PS order parameter $\mathcal O_{\rm F}$
and compare with the behavior of empty PS order parameter,
$\mathcal O_{\rm E}$, which is discussed in the main text. Note
that their definitions are indicated in Fig.~\ref{sm_susc}(a).

\begin{figure}[h!]
  \includegraphics*[width=0.9\linewidth]{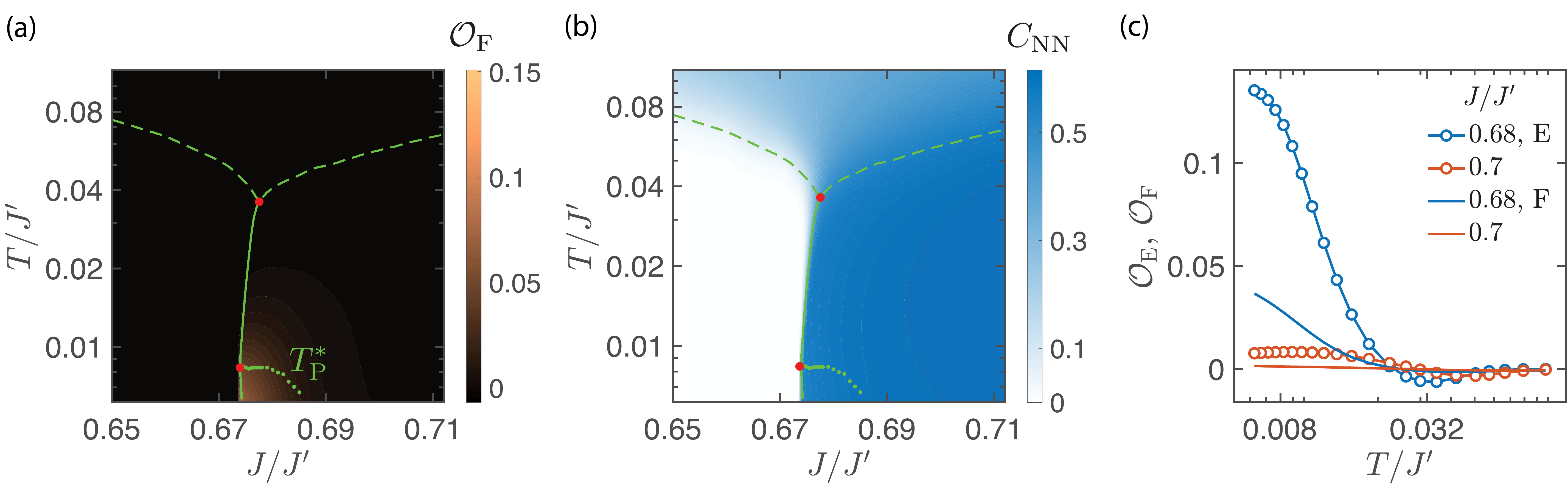}
  \caption{Contour plots of (a) full PS order parameter $\mathcal{O}_F$,
  and (b) inter-dimer correlator $C_{\rm NN}$. Scale of the colorbar in (a)
  is the same as that used in Fig.~2(b) of the main text.
  Green dashed, solid and {dotted lines} are peaks of $C_m/T$, 
  two red points are CP and CEP, respectively [c.f.
  Fig.~1(a)]. (c) Temperature dependence of order parameters $\mathcal{O}_E,
  \mathcal{O}_F$ for both empty and full PS states, respectively.
  }
  \label{smFig:z2op}
\end{figure}

Before presenting numerical results, we firstly explain how to obtain these
order parameters in practice. Due to the underlying snake path geometry
for the MPO used, horizontal bonds have a tendency to be
weaker than the vertical bonds. As illustrated in Fig.~1(c) of the main text,
this is due to the fact that vertical bonds are long-ranged ones while horizontal bonds are usually
nearest neighboring. Moreover, due to the cylindrical geometry used, boundary effects can induce a particular empty PS pattern
near the boundaries~\cite{yang2022}. In practice, we sum both two horizontal
bonds and two vertical bonds emanating from each of the four sublattices of
the \emph{central} unit cell, with appropriate signs taken into account as
illustrated in the inset of Fig.~\ref{sm_susc}(a).

In Fig.~\ref{smFig:z2op}(a), we show the corresponding contour plot of the $\ZZ$ 
symmetry breaking order parameter for full PS, and compare the results to Fig.~2(b)
in the main text. It clearly shows that $\mathcal O_{\rm F}$ remains zero until the 
temperature deceases also to $T^\ast_{\rm P}$. Moreover, below this temperature 
scale, $\mathcal O_{\rm F}$ is much smaller than $\mathcal O_{\rm E}$, which 
indicates that the empty plaquette indeed wins the competition in the $\ZZ$ 
symmetry breaking order. In Fig.~\ref{smFig:z2op}(c), we compare the temperature 
evolution of $\mathcal O_{E}$ and $\mathcal O_{\rm F}$, and find indeed the former 
is larger than the latter at low temperature in the PS phase. On the other hand, in the 
AF phase both parameters remain small till the lowest temperature.

As a sidenote, we also present the contour plot of the inter-dimer correlator
$C_{\rm NN}$ in Fig.~\ref{smFig:z2op}(b), which shows that it becomes 
nonzero outside the DS phase. This complementary behaviors in comparison 
with the intra-dimer correlator $C_{\rm D}$ given in Fig.~2(a) of the main text
confirm the nature of the first-order DS-PS phase transition.

\subsection{Plaquette singlet susceptibility}

\begin{figure}[h!]
\centering
\includegraphics[width=0.48\linewidth]{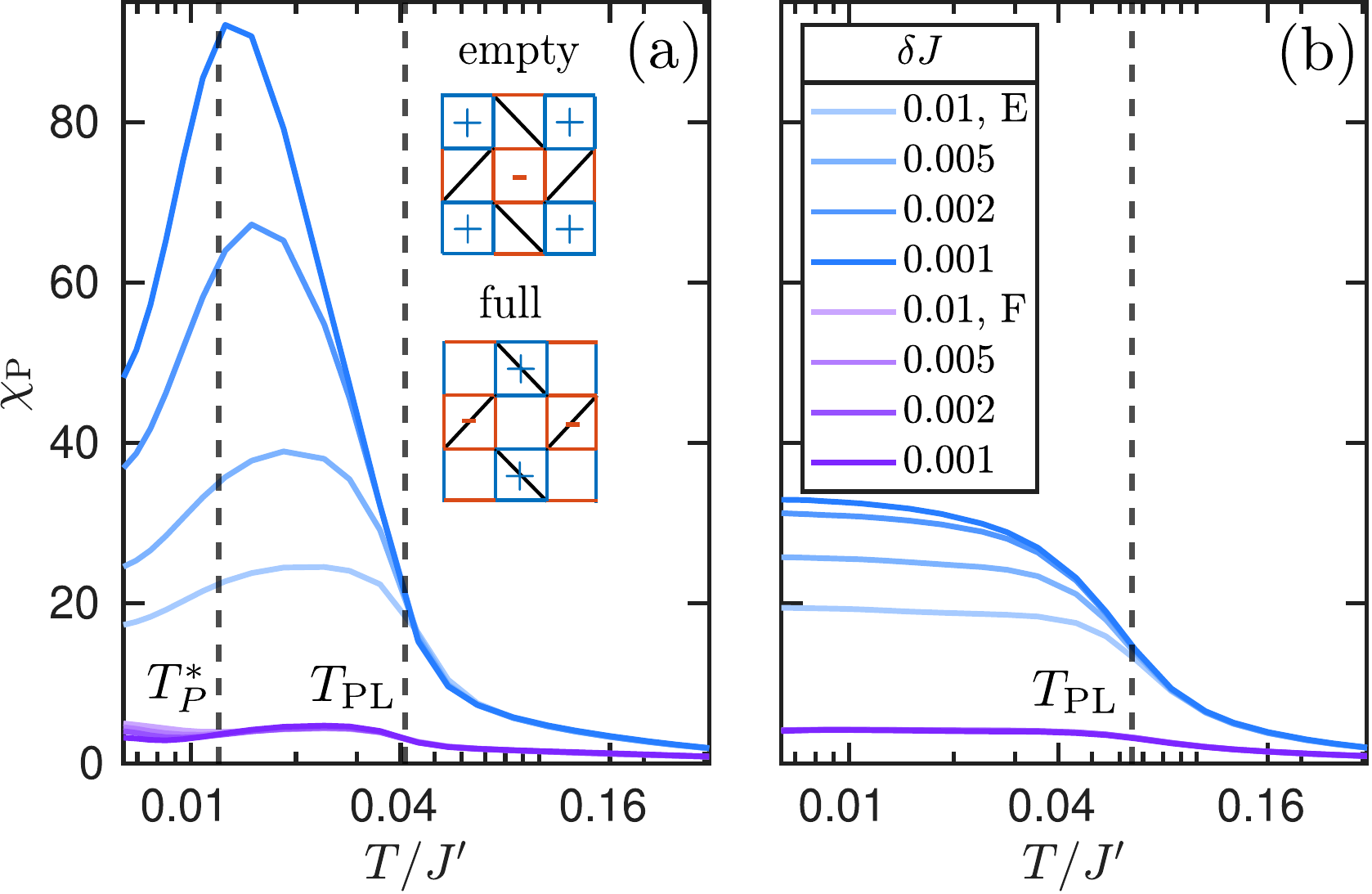}
\caption{PS susceptibility $\chi_{\rm P}$ at (a) $J/J'=0.68$ and (b) $J/J'=0.71$, 
for empty ($P = {\rm E}$) and full ($P = {\rm F}$) PS order parameters, and 
illustrated in inset of the left panel. Four different values of $\delta J$ are 
used to ensure convergence of $\chi_{\rm P}$ calculations.
}
\label{sm_susc}
\end{figure}

To resolve the issue that whether the $\ZZ$ symmetry breaking occurs
among the empty or full plaquettes, here we reveal their competitions
across the full temperature windows. Namely, we calculate PS susceptibility
for both empty and full PS orders, defined by $ \chi_{\rm P} = {[\mathcal 
O_{\rm P}(\delta J_{\rm P}) - \mathcal O_{\rm P}(0)]}/{\delta J_{\rm P}}$,
with $\rm{P} = \rm{E\text{ or }F}$. Here $\delta J_{\rm P}$ means
a small ``pinning field'' for inter-dimer coupling $J\rightarrow J\pm 
\delta J_{\rm P}$, as shown in the inset of Fig.~\ref{sm_susc}(a). 
In Fig.~\ref{sm_susc}(a), we present two types of PS susceptibility 
for $\alpha=0.68$. It clearly shows that $\chi_{\rm E}$ is larger than 
$\chi_{\rm F}$ even for relatively high temperatures. Moreover, the 
former increases much faster than the latter in the PL regime below 
$T_{\rm PL}$, and reaches its maximal value around $T^\ast_{\rm P}$,
indicating the occurrence of $\ZZ$ symmetry breaking there. Therefore, 
we conclude that the empty PS instability predominates over the full one 
in the low-temperature regime in the temperature-pressure phase diagram. 
Moreover, although the full PS susceptibility is relatively small, it is nonzero 
and indicates the existence of a competition that may be responsible for 
the very low PS order transition temperature. Symmetry breaking eventually 
occurs at $T^\ast_{\rm P}$, which is two orders of magnitude smaller than 
the coupling strength of the model, a manifestation of PS order competition 
and strong spin fluctuations.

In Fig.~\ref{sm_susc}(b), the same quantities are plotted for $\alpha=0.71$. 
We find again that $\chi_{\rm E}$ increases much faster than
$\chi_{\rm F}$ for $T \lesssim T_{\rm PL}$. However, this increase soon saturates
to a constant value as the temperature further decreases. It indicates that the
$\ZZ$ symmetry breaking order has never been built up for this case, and the
two types of PS correlations remain fluctuating. Indeed, the ground state for
 this case is the AF N\'eel state as obtained by DMRG
(c.f. Sec.~\ref{smsec2}).

\subsection{{Pressure-induced phase transitions} in the SS model}
In this subsection, we provide more evidence on the DS-PS (and DS
to the plaquette-single liquid, PL) phase transitions, and in the end also 
discuss the mysterious PS-AF phase
transition from our finite-$T$ data. Note the intra-dimer correlator,
$C_{\rm D}$, is used in the main text to reveal the existence of the
first-order transition line.

\begin{figure}[t!]
\includegraphics*[width=0.85\linewidth]{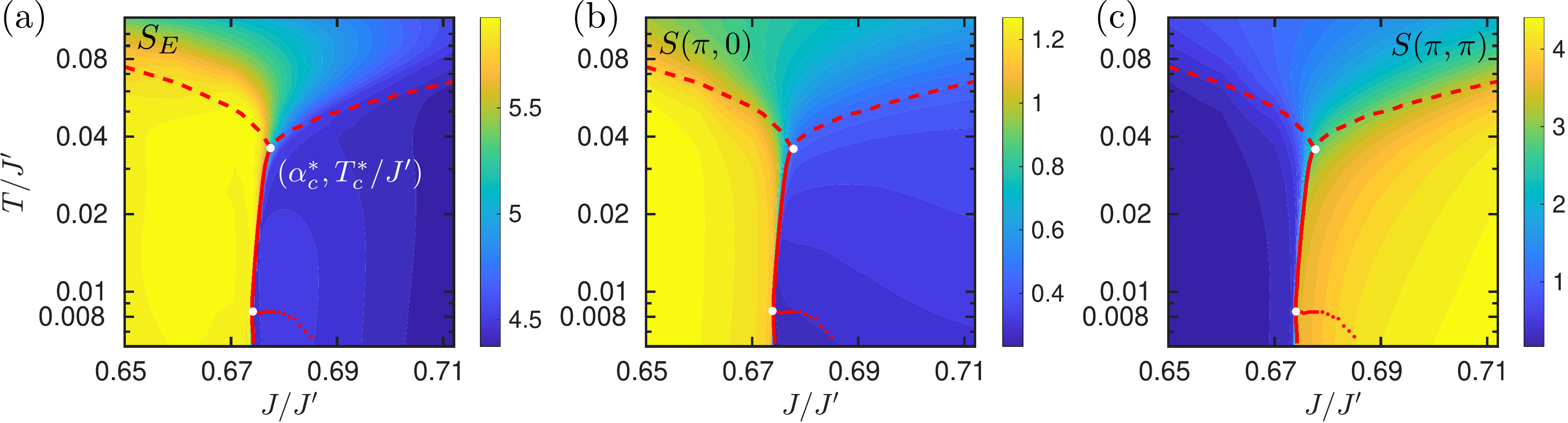}
\caption{Contour plots of (a) the MPO EE $S_E$, (b) static spin structure 
factor $S(\pi,0)$ and (c) $S(\pi,\pi)$. The red dashed, solid and dotted lines 
are determined from the peaks of $C_m/T$, and the two white dots represent 
the CP and CEP, respectively.} 
\label{smFig:1stpt}
\end{figure}

\textit{MPO entanglement entropy.---}
Taking an MPO as a supervector, a Schmidt decomposition of this purified
wavefunction can be performed and the entanglement entropy (EE) $S_E$
between the two parts of the system can be computed. Here we divided the
density matrix MPO at the middle bond and compute the corresponding MPO
EE in Fig.~\ref{smFig:1stpt}. It shows that the DS phase has larger values
than other phases, and there is a jump at the first-order transition line. At
higher temperatures, MPO entanglement entropy instead exhibits a smooth
crossover as tuning $J/J'$.

\textit{Static spin structure factor.---}
Another interesting quantity to characterize the spin states and transitions
is the static spin structure factor $S(\vb k) = N^{-1} \sum_{i,j} e^{-i\vb k \cdot
(\vb r_i-\vb r_j)} \langle \vb S_i \cdot \vb S_j \rangle$, where $N$ is the total
lattice sites {and $\vb k = [k_x, k_y]$}. A particular temperature evolution of 
this quantity is given in the second column of Fig.~1(c) for $J/J'=0.676$. 
Here we focus on the two particular momentum points, $\vb k =(\pi,0)$ and 
$\vb k = (\pi,\pi)$. The DS phase has a structure peak at $(\pi, 0)$, while the
other phases of this model exhibits peak at $(\pi,\pi)$ \cite{chung2001}.
Contour plots of these two quantities are shown in Fig.~\ref{smFig:1stpt}(b)
and (c). Indeed in Fig.~\ref{smFig:1stpt}(b) we see that in the DS phase,
the $S(\pi, 0)$ takes the larger values than those of PL, PS, and AF phases;
while the case is reversed in Fig.~\ref{smFig:1stpt}(c). The first-order line
again is evident by a jump for $T<T_{c}^\ast$ as $J/J'$ varies; while they
can changes smoothly for higher temperatures above $T_{c}^\ast$.

\textit{On the PS-AF transition.---}
According to the ground-state DMRG calculations, a phase possesses
long-range AF order appears for high pressure (c.f., Sec.~\ref{smsec2}).
It follows that there must exist a QPT between the empty PS and AF phases 
in the SS model. From the specific heat and correlation results in Figs.~1 
and 2 of the main text, a typical first-order transition is ruled out and they 
point to a possible existence of QCP there. However, all quantities calculated
at finite temperature in this work show no clear evidence for the putative
PS-AF phase transition. This can be due to the fact that this transition is
a very weak one as proposed in Ref.~\cite{corboz2013,xi2023} and the
relevant temperature scale is extremely low and out of reach from our
numerics. At present we are not able to determine the nature of this QPT, 
nevertheless, our study indicates that this phase transition may be equally 
hard to detect experimentally even down to low temperature.

\subsection{Critical scaling in order parameter $C_{\rm D}$}
As shown in Fig.~2(a) of the main text, the intra-dimer correlator $C_{\rm D}
\equiv -\expval{\vb S_i \cdot \vb S_j}_{\rm D}$, for sites $i$ and $j$ belonging
to a diagonal dimer $J'$ bond, serves as the order parameter detecting the
first-order transition of DS-PL. Namely, for the DS (PL) phase, $C_{\rm D}$
is positive (negative). In this sense, it plays the role of density-type order 
parameter as in water's phase diagram for distinguishing the liquid and gas 
phases. Since the difference is defined not by a change in symmetry, but by 
a scalar which takes two different values at the discontinuity, the corresponding 
critical point belongs to the 2D Ising universality class. Here we confirm this 
argument by studying the critical scaling behavior of $C_{\rm D}$.

\begin{figure}[h!]
\includegraphics*[width=0.85\linewidth]{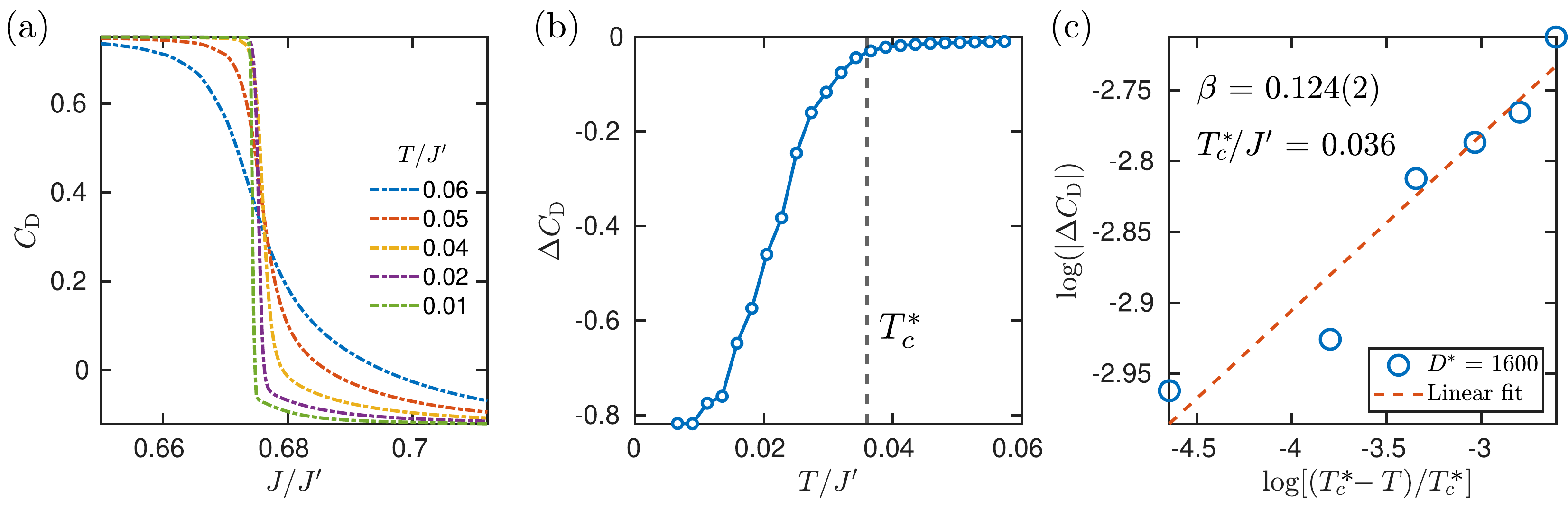}
\caption{(a) Dimer correlator $C_{\rm D}$ as a function of $J/J'$.
(b) Discontinuity of dimer correlator, $\Delta C_{\rm D}$, at the
first-order transition line, is plotted as a function of temperature.
(c) Log-log plot of the $\Delta C_{\rm D}$ v.s. the reduced temperature,
from which an exponent of $\beta \approx 1/8$ can be extracted that
agrees with 2D Ising universality class.}
\label{smFig:dimercritical}
\end{figure}

In Fig.~\ref{smFig:dimercritical}(a), we show $C_{\rm D}$ as a function of
$J/J'$ for several different temperatures. It is found that there is indeed an
abrupt jump in $C_{\rm D}$ for low temperatures, while the change becomes 
smooth at higher temperatures. We extract this jump in $C_{\rm D}$ at the 
first-order transition line, $\Delta C_{\rm D}$, show the results in Fig.
\ref{smFig:dimercritical}(b) as a function of $T$. We find 
the absolute value of $\Delta C_{\rm D}$ 
decreases monotonically as the temperature increases and vanishes when
approaching $T_c^\ast$.

Location of the critical point is determined as the crossing point between
three lines determined from the peaks of $C_m/T$ as shown in Fig.~1(a),
which gives $(\alpha_c^\ast, T^\ast_c /J') = (0.678, 0.036)$. We then
define the reduced temperature $\tilde{t} = (T_c^\ast-T)/T_c^\ast$, and the
discontinuity $\Delta C_{\rm D}$ as a function of $\tilde{t}$ is shown in
Fig.~\ref{smFig:dimercritical}(c) in a log-log scale. Through a linear fit,
we extract the critical exponent $\beta$ from $|\Delta C_{\rm D}| =
\tilde{t}^\beta$, which leads to $\beta \simeq 0.124$, in agreement
with the 2D Ising exponent $\beta = 1/8$.

\subsection{Temperature derivatives of the intra- and inter-dimer singlet correlators}
Two types of local correlations, intra-dimer $C_{\rm D}$ and inter-dimer
$C_{\rm NN}$, play a central role in characterizing the finite-$T$ phases
of the SS model. As shown by Fig.~2(a) in the main text and
Fig.~\ref{smFig:z2op}(b) above,
$C_{\rm D}$ ($C_{\rm NN}$) dominates in the DS (PS) phase, respectively.
It is thus important to further investigate their temperature evolution.
In particular, one straightforward quantity is their partial derivative
with respect to temperature $T$, which has a clear physical meaning:
$\gamma_P = -(\partial S/\partial P)_T \sim \partial(\mathcal{A} \,
C_{\rm D} + \mathcal{B} \, C_{\rm NN})/\partial T$ (with $\mathcal{A},
\mathcal{B}$ model-dependent coefficients) characterizes the
magnetic barocaloric effect.

\begin{figure}[h!]
\includegraphics*[width=0.6\linewidth]{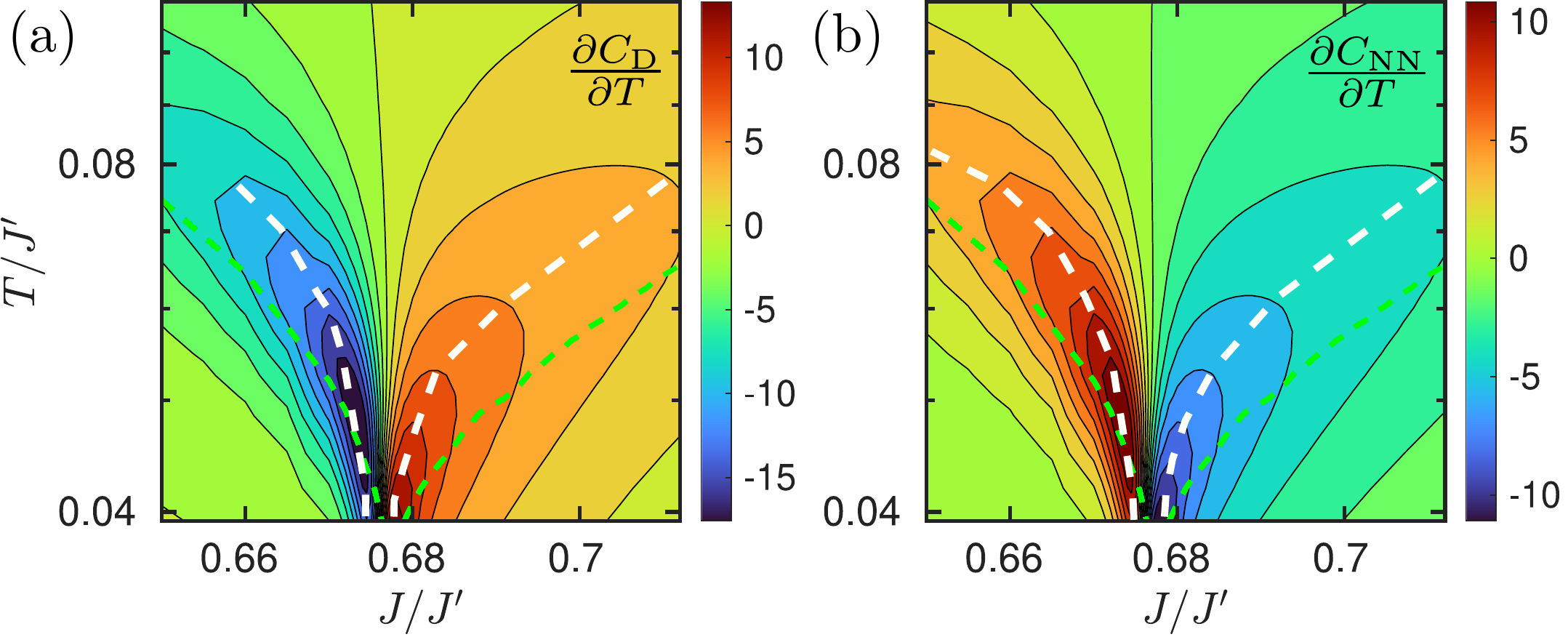}
\caption{{Partial derivative of (a) intra-dimer correlator $C_{\rm D}$ and
(b) inter-dimer correlator $C_{\rm NN}$ with respect to temperature $T$
(under fixed pressure $P$). The green dashed lines are peaks of $C_m/T$
[c.f. Fig.~1(a) of the main text]. The white dashed lines connects the
extrema (peaks or dips) of the contour lines.}}
\label{sm_pcd_pcnn}
\end{figure}

In Fig.~\ref{sm_pcd_pcnn}(a) and (b), we show the temperature derivatives
of these two local correlators. It is found that near the critical point there are
two white dashed lines forming local extrema, analogous to the peaks of
$C_m/T$ (indicated by the green dashed lines). The derivatives are found
to have different sign in the DS and PL/PS regimes, and the peak values
increase as temperature decrease.

\section{More results on the SS model in finite magnetic fields}

\subsection{Magnetocaloric evidence for the spin supersolid transition}
In the main text, we find the magnetic specific heat $C_m/T$ exhibits
certain signatures of the PS-SSS QPT. Here we present more clear 
evidences in terms of magnetocaloric responses.

\begin{figure}[h!]
\includegraphics*[width=0.65\linewidth]{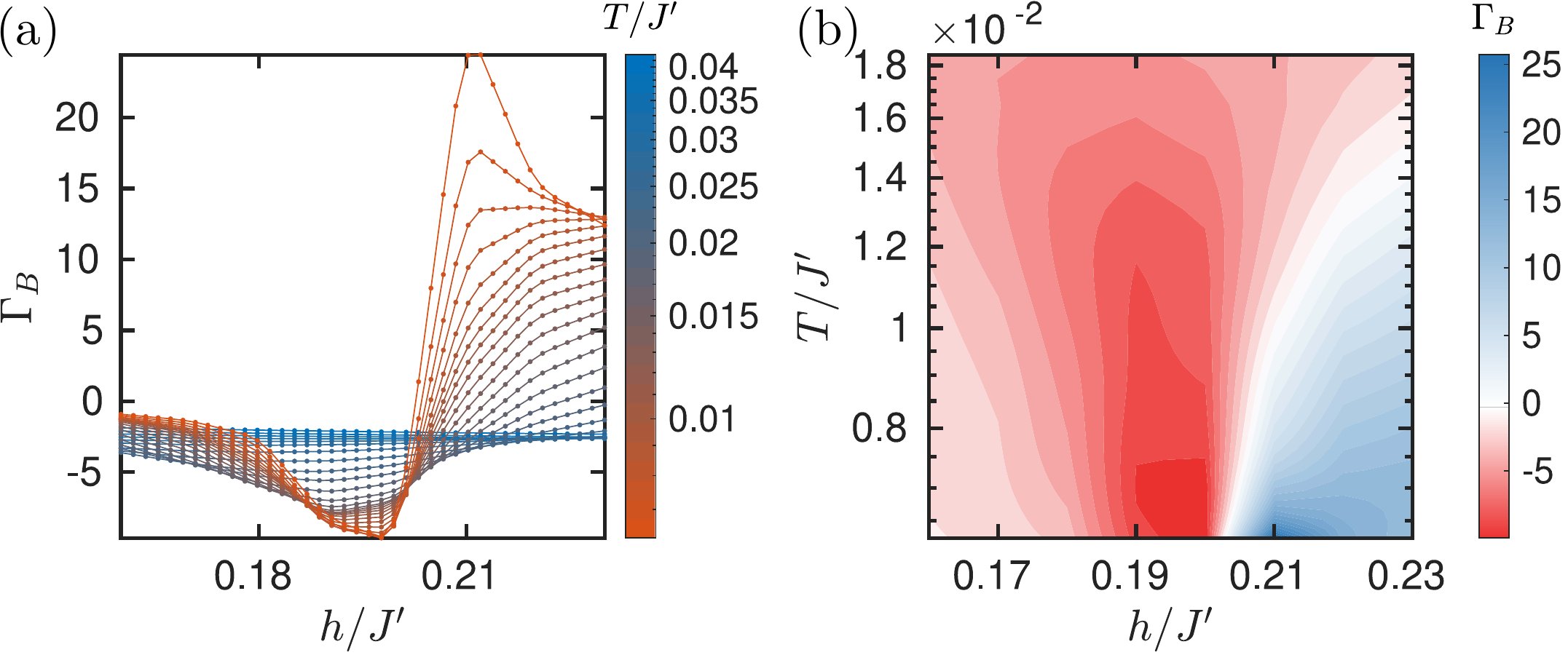}
\caption{(a) Magnetic {\gr} $\Gamma_B$ as a function of $h/J'$ for
different low temperatures, and (b) the corresponding contour plot.
The results are obtained for $J/J' = 0.68$ with $D=3200$ {U(1)} states
retained in the simulations.}
\label{smFig:entropy}
\end{figure}

In Fig.~\ref{smFig:entropy}, we show the computed magnetic {\gr},
defined by $\Gamma_B = -\frac{1}{T} \frac{(\partial S/\partial h)_T}
{(\partial S/\partial T)_h}$, as a function of the external magnetic field
$h/J'$, for different temperatures. The thermal entropy results are
obtained with $D=3200$ {U(1) states}. It is found that $\Gamma_B$ 
indeed tends to diverge near the QPT with a sign change at low 
$T$~\cite{zhu2003,garst2005}, which are shown in the $\Gamma_B$ 
vs. fields $h$ in Fig.~\ref{smFig:entropy}(a) and the corresponding 
contour plot in Fig.~\ref{smFig:entropy}(b). We note that this prominent 
magnetocaloric response can be used to probe the SSS transition 
experimentally from magnetocaloric measurements.

\subsection{Temperature evolution of the spin structure factors}

\begin{figure}[h!]
\includegraphics*[width=\linewidth]{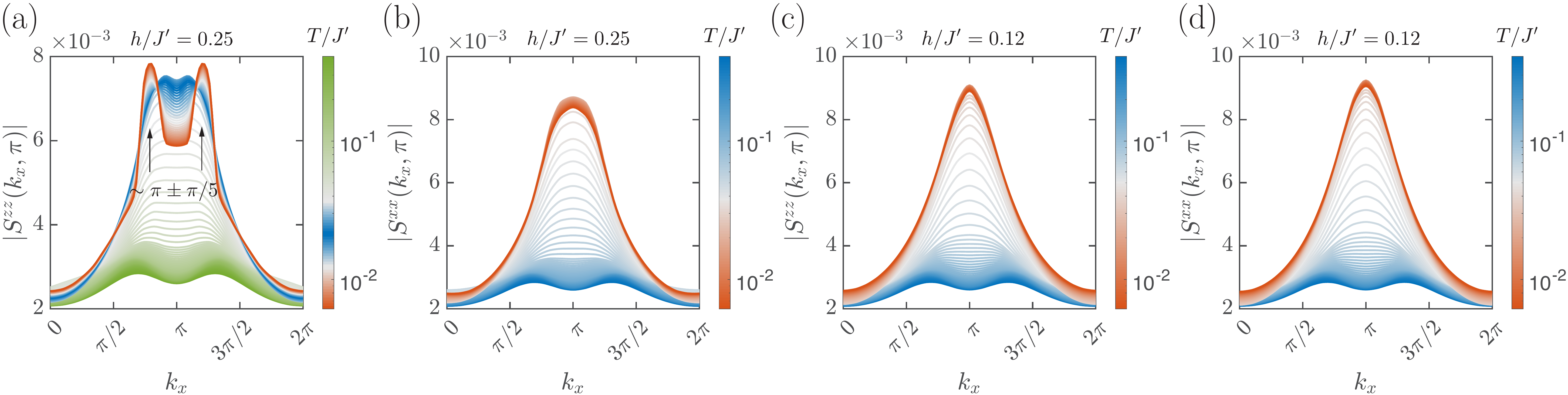}
\caption{Spin structure factors {$S^{zz}(k_x, k_y)$} (a,c) 
and {$S^{xx}(k_x, k_y)$} (b,d), plotted along the lines with 
fixed $k_y=\pi$.}
\label{smFig:ssfxxzz}
\end{figure}

Here we discuss nature of the state sitting on the right bottom corner 
of Fig.~4(a) in the main text, where a new phase different {from} the
high-temperature PL phase is expected. As revealed by the DMRG
calculations elaborated in Sec.~\ref{smsec2}, the ground state 
corresponds to an intriguing $10\times2$ stripy SSS phase, where both 
the discrete translational symmetry (diagonal ``solid'' order) and the spin 
rotation symmetry along $z$ axis (off-diagonal ``superfluid'' order) are 
spontaneously broken. As temperature ramps up, there are in principle
three possibilities for the broken symmetries to restore: either the discrete 
or continuous symmetry firstly restores, or they occur at the same temperature 
(which is not very likely though).

To verify which scenario happens in the SS model for $J/J' = 0.68$, we
consider the spin structure factors in the spin $z$ and $x$ components
defined as
\begin{equation}
S^{\gamma\gamma}(\vb k) =\frac{1}{N} \sum_{i,j} e^{i\vb k \cdot (\vb r_i - \vb r_j)}
\expval{S^\gamma_i S^\gamma_j},\qfor \gamma = x,y,z.
\end{equation}
In the PS phase, both $S^{xx}(\vb k)$ and $S^{zz}(\vb k)$ should
peak at $\vb k = M = (\pi,\pi)$~\cite{chung2001}; while in the SSS
phase, we expect that both of them exhibit peaks at $(\pi\pm\pi/5,\pi)$,
since the discrete symmetry breaking pattern is $10\times 2$,
as illustrated explicitly in Fig.~4(c) of the main text.

In Fig.~\ref{smFig:ssfxxzz}, we fix $k_y=\pi$ and show
the temperature evolutions of spin structure factors. It is found that for
$h/J'=0.25$, the peak of $S^{zz}$ firstly develops around $M$ for temperature
reaching $T_{\rm PL}$, then it shifts to other two symmetric momenta
as temperature further lowers, suggesting a discrete symmetry breaking 
of the system in this temperature regime [c.f. Fig.~\ref{smFig:ssfxxzz}(a)].
On the other hand, the peak of $S^{xx}$ remains at $M$ down to the
lowest accessible temperature [c.f. Fig.~\ref{smFig:ssfxxzz}(b)]. It is likely that the BKT temperature is 
with an even lower value out of reach by our finite-temperature calculations.
As a comparison, we also show the case with $h/J'=0.12$ (in the PS phase), 
where it is found that both components establish clear peaks at $M$ without 
any splitting, for temperatures below $T_{\rm PL}$ [c.f. Fig.~\ref{smFig:ssfxxzz}(c-d)]. We also note that this peak 
is sharper for $h/J'=0.12$ than those of $h/J'=0.25$ case.
We note that these temperature evolution of spin structure factors
can be experimentally detected via neutron scattering experiments, and
provide experimental evidence for the spin supersolid phase in {\scbo}
under finite magnetic fields.


\end{document}